\documentclass[twocolumn,prb,amsmath,amssymb,floatfix]{revtex4}

\usepackage{bbold}
\usepackage{bm}
\usepackage{color}
\usepackage{dcolumn}
\usepackage{feynmf} 
\usepackage{graphics}
\usepackage{graphicx}
\usepackage{subfigure}
\usepackage{slashed}
\usepackage{tensor}
\usepackage{placeins}
\usepackage{natbib}

\usepackage[colorlinks=true,urlcolor=black]{hyperref}

\hypersetup{backref = true,pagebackref = true,hyperindex = true, colorlinks = true,
  breaklinks = true, urlcolor = blue, linkcolor = blue, bookmarks = true,
  bookmarksopen = true, citecolor=red}

\def\Tr{\mbox{Tr}}

\def\al{\alpha}

\def\om{\omega}
\def\veps{\varepsilon}

\def\be{\begin{equation}}
\def\ee{\end{equation}}
\def\bea{\begin{eqnarray}}
\def\eea{\end{eqnarray}}
\def\bse{\begin{subequations}}
\def\ese{\end{subequations}}
\def\bc{\begin{center}}
\def\ec{\end{center}}

\def\ra{\rightarrow}

\def\nonum{\nonumber}

\def\I{{\rm i}}

\def\D{{\rm d}}
\def\Ord{{\rm O}}

\def\Sp{{\slashed p}}

\def\Sk{{\slashed k}}

\newcommand{\ie}{{\it i.e.}}
\newcommand{\eg}{{\it e.g.}}

\newcommand{\comment}[1]{}

\catcode`,\active

\catcode`\,12

\allowdisplaybreaks

\begin{document}

\title{Field theoretic renormalization study of reduced quantum electrodynamics \\ and applications to the ultra-relativistic limit of Dirac liquids}

\author{S.~Teber$^1$ and A.~V.~Kotikov$^2$}
\affiliation{
$^1$Sorbonne Universit\'e, CNRS, Laboratoire de Physique Th\'eorique et Hautes Energies, LPTHE, F-75005 Paris, France.\\
$^2$Bogoliubov Laboratory of Theoretical Physics, Joint Institute for Nuclear Research, 141980 Dubna, Russia.}

\date{\today}

\begin{abstract}
The field theoretic renormalization study of reduced quantum electrodynamics (QED) is performed up to two loops. 
In the condensed matter context, reduced QED constitutes a very natural effective relativistic field theory describing (planar) Dirac
liquids, \eg, graphene and graphene-like materials, the surface states of some topological insulators and 
possibly  half-filled fractional quantum Hall systems. From the field theory point of view, the model involves an effective (reduced) gauge field
propagating with a fractional power of the d'Alembertian in marked contrast with usual QEDs. The use of the Bogoliubov-Parasiuk-Hepp-Zimmermann prescription allows
for a simple and clear understanding of the structure of the model. In particular, in relation with the ultra-relativistic limit of graphene, 
we straightforwardly recover the results for both the interaction correction to the optical conductivity: 
$\mathcal{C}^*=(92-9\pi^2)/(18\pi)$ and the anomalous dimension
of the fermion field:  $\gamma_{\psi}(\bar{\al},\xi) = 2 \bar{\al}\,(1-3\xi)/3 -16\,%
\left( \zeta_2 N_F + 4/27 \right)\, \bar{\al}^2 + \Ord(\bar{\al}^3)$, where $\bar{\al} = e^2/(4\pi)^2$ and $\xi$ is the
gauge-fixing parameter.
\end{abstract}

\maketitle

\begin{fmffile}{fmfrqed-bphz}

\section{Introduction}

It is well-known that Dirac liquids posses an infra-red (IR) Lorentz-invariant
fixed point [\onlinecite{Gonzalez:1993uz}]. 
This fact is actually generic to systems with stable Fermi (or Dirac) points, see the textbook [\onlinecite{Volovik2009book}]. 
In these systems, low-energy excitations have a gapless linear, relativistic-like, spectrum as known theoretically 
for a long time in graphene [\onlinecite{PhysRev.71.622,Semenoff:1984dq}] and in the so-called
spin-liquid phases of high temperature superconductors [\onlinecite{Marston:1989zz,Ioffe:1989zz}]. Such low-energy Dirac-like
energy dispersions are by now well observed experimentally in various materials which are under active study, \eg, graphene [\onlinecite{Novoselov:2005kj}],
(artificial) graphene-like materials [\onlinecite{polini2013artificial}], surface states of topological
insulators [\onlinecite{RevModPhys.82.3045}], the so-called Dirac-Weyl materials 
[\onlinecite{liu2014dirac3d,neupane2013dirac3d,liu2013dirac3d,PhysRevLett.113.027603,huang2014weyl,xu2015weyl}]
see the review [\onlinecite{Wehling2014Dirac}] on these three-dimensional analogues of graphene and, very recently,  half-filled fractional 
	quantum Hall systems [\onlinecite{PanCF:2017}].

The IR Lorentz invariant fixed point arises from the long-range Coulomb interaction among the Dirac fermions which
enforces the flow of the Fermi velocity, \eg, $v \approx c/300$ at experimentally accessible scales for graphene, to the velocity of light, $c$, in the IR with a corresponding flow of 
the fine structure constant, \eg, $\al_g \approx e^2/4\pi \veps \hbar v \approx 2.2$ for graphene, to the usual fine structure constant, $\al \approx 1/137$, in the IR.
Moreover, in the case of planar systems such as graphene, the electrons are confined to a three-dimensional space-time, $d_e = 2+1$, while 
interactions between them are mediated by four-dimensional photons, $d_\gamma = 3+1$.
The Lorentz invariant fixed point may therefore be effectively described by a massless relativistic quantum field theory (QFT)
model whereby $d_e$-dimensional fermions interact via a $d_\gamma$-dimensional $U(1)$ gauge-field.
Such a model belongs to the class of reduced quantum electrodynamics (QED) [\onlinecite{Gorbar:2001qt}], reduced QED$_{d_\gamma,d_e}$ or simply QED$_{d_\gamma,d_e}$, also known as
pseudo QED [\onlinecite{Marino:1992xi}] and even more recently as mixed-dimensional QED [\onlinecite{Son:2015xqa}].
Notice that in the particular case where gauge and fermion fields live in the same space-time, $d_\gamma=d_e=d$, reduced QEDs correspond to usual QED$_d$ while in the reduced case $d_e < d_\gamma$.
Early motivations for the study of reduced theories came from interest in branes [\onlinecite{Gorbar:2001qt,Dimopoulos:2000iq}], 
dynamical chiral symmetry breaking on a brane [\onlinecite{Gorbar:2001qt}], conformal field theory [\onlinecite{Kaplan:2009kr}] (and reference therein),
as well as potential applications to condensed matter physics systems in relation with the quantum Hall effect and high temperature superconductivity [\onlinecite{Marino:1992xi,Dorey:1991kp,Kovner:1990zz}].
In Ref.~[\onlinecite{Teber:2012de}], QED$_{d_\gamma,d_e}$ was advocated as a minimal model to study the infra-red Lorentz invariant fixed point 
of Dirac liquids with a special focus on QED$_{4,3}$ relevant to intrinsic (or undoped) disorder-free graphene and similar planar materials.
In the last years, there has been an increasing number of studies focusing on reduced QED and in particular QED$_{4,3}$ in relation with, {\it e.g.},
transport and spectral properties [\onlinecite{Kotikov:2013kcl,Herbut:2013kca,Kotikov:2013eha,Teber:2014hna}], see also the short review [\onlinecite{Teber:2016unz}], optical
properties [\onlinecite{Valenzuela:2014uia,Hernandez-Ortiz:2015wua}], quantum Hall effect [\onlinecite{Marino:2015uda,Kooi:2017ugi,Son:2015xqa}]
and dynamical chiral symmetry breaking [\onlinecite{Kotikov:2016yrn,Alves:2017fij}] in planar systems.
Moreover, QED$_{4,3}$ was shown to be unitary [\onlinecite{Marino:2014oba}], its properties
under the Landau-Khalatnikov-Frandkin transformation were studied [\onlinecite{Ahmad:2016dsb}], its precise relation to QED$_3$ understood [\onlinecite{Kotikov:2016yrn}], 
it was shown to possess a strong-weak duality mapping the coupling constant $e$ to $\tilde{e}=8\pi/e$ with a self-dual point at $e^2=8\pi$ (or $\al =2$) [\onlinecite{Hsiao:2017lch}] 
and, even more recently, it has been studied as an interacting boundary conformal field theory [\onlinecite{Herzog:2017xha}].

Though essentially of academic interest, a thorough understanding of the IR fixed point is a necessary prerequisite to set on a firm ground the study of 
the physics away from the fixed point which is closer to the experimental reality but more difficult to study theoretically. Indeed, in the non-relativistic 
limit there is often no definitive agreement on the precise value of important quantities directly related to interaction effects;
in relation with graphene, let's for example mention two quantities that have been the subject of extensive work during the last decade: the value of the interaction correction to the optical conductivity, see, \eg, Refs.~[\onlinecite{Herbut08.PhysRevLett.100.046403,Mishchenko2008,%
        Juricic:2010dm,Sheehy09.PhysRevB.80.193411,Abedinpour11.PhysRevB.84.045429,Sodemann12.PhysRevB.86.115408,%
        Gazzola13.0295-5075-104-2-27002,Rosenstein13.PhysRevLett.110.066602,%
        Teber:2014ita,Link16.PhysRevB.93.235447,Boyda:2016emg,Stauber17.PhysRevLett.118.266801}]
and the value of the critical coupling constant for dynamical gap generation, see, \eg, 
	Refs.~[\onlinecite{Khveshchenko:2001zz,Gorbar:2002iw,Leal:2003sg,Son07.PhysRevB.75.235423,VafekCase08,Khveshchenko:2008ye,Liu09.PhysRevB.79.205429, Gamayun:2009,Drut:2008rg,Drut:2009aj,Drut:2009zi,WangLiu11a,WangLiu11b,Gonzalez12.PhysRevB.85.085420,Wang2012,Buividovich12.PhysRevB.86.245117,Ulybyshev:2013swa,Popovic13.PhysRevB.88.205429,Gonzalez15.PhysRevB.92.125115,Katanin16.PhysRevB.93.035132,Carrington:2017hlc}].
It turns out that QED$_{4,3}$ is an ideal playground to compute both of these quantities, 
see [\onlinecite{Teber:2012de,Kotikov:2013kcl,Kotikov:2013eha}] (as well as [\onlinecite{Teber:2016unz}] for a short review) and [\onlinecite{Kotikov:2016yrn}] respectively.
The reason is that all the powerful multi-loop machinery originally developed in particle physics and statistical mechanics to compute (massless) Feynman diagrams, see, \eg,
Refs.~[\onlinecite{Kazakov:1984km,Kotikov:1995cw}] and also the lectures [\onlinecite{Kazakov:1984bw}], can be applied to reduced QED in order
to rigorously understand the perturbative structure of the model as well as some of its non-perturbative features.
Interestingly, the odd dimensionality of space-time together with the (related) presence of Feynman diagrams with non-integer indices brings a lot of novelties 
(as well as highly non-trivial additional complications) with respect to what is usually known from the study of $(3+1)$-dimensional theories, see 
Refs.~[\onlinecite{Kotikov:2013kcl,Kotikov:2013eha,Teber:2016unz,Kotikov:2016rgs}] for a systematic computation of non-trivial master integrals in QED$_{4,3}$ up to two-loops.
Besides loop calculations, a non-trivial aspect of reduced QED is related to the peculiar structure of its (sub-)divergent graphs.
It is then the purpose of renormalization to give a prescription on how to deal with these (sub-)divergences. In all previous references, the
so-called conventional renormalization has been used to achieve this purpose.

In this paper, we will focus on the field theoretic renormalization study of reduced quantum electrodynamics and in particular QED$_{4,3}$. We 
will assume that all the needed  master integrals are known and proceed in renormalizing the model and extracting anomalous dimensions and renormalized correlation functions with the 
help of the recursive subtraction scheme, the so-called $R$-operation, of Bogoliubov and Parasiuk~\cite{Bogoliubov:1957gp} and
Hepp~\cite{Hepp:1966eg} or its solution known as Zimmermann's forest formula~\cite{Zimmermann1969}, see also the textbook 
Ref.~[\onlinecite{Collins:1984xc}].~\footnote{Let's also note the more recent Hopf algebraic formulation~\cite{Kreimer:1997dp} of renormalization (see Ref.~[\onlinecite{Panzer:2014kia}] for a 
recent review). Its application to our model is beyond the scope of our present study.} 
 The power of the BPHZ prescription, with respect to conventional renormalization, lies in the fact that it applies diagram by diagram, gives a very clear and 
unambiguous prescription on how to subtract (sub-)divergences and is conveniently automated.~\cite{Batkovich:2014rka,Herzog:2017bjx} We will apply such a prescription to computing the
interaction correction to the optical conductivity and the anomalous dimension of the fermion field (which plays an important role
with respect to the critical coupling constant for dynamical gap generation). As will be shown in detail in the following, 
the obtained results are in complete agreement with those obtained via conventional renormalization thereby lifting any possible ambiguity as to their
value at the IR fixed point.

The paper is organized as follows. In Sec.~\ref{Sec:RQED}, we motivate the study of reduced QED and set up the general notations and conventions.
In Sec.~\ref{Sec:One-Loop}, we recall the one-loop structure of the model. In Secs.~\ref{Sec:Two-Loop-Pi} and \ref{Sec:Two-Loop-Sigma}, we then 
focus on the renormalization of the polarization operator and the fermion self-energy, respectively. We conclude in Sec.~\ref{Sec:Conclusion} and define some basic master integrals
appearing in the text in App.~\ref{App:MI}. In the following, we work in units where $\hbar=c=1$.

\section{General approach and model}
\label{Sec:RQED}

\subsection{General approach}

The most general low-energy effective action (model I) describing a disorder-free intrinsic Dirac liquid reads (in Minkowski space):
\begin{flalign}
	S &= \int \D t\, \D^{D_e} x\, \left[ \bar{\psi}_\sigma \left( \I \gamma^0 \partial_t + \I v \vec{\gamma} \cdot \vec{\nabla}\,\right) \psi^\sigma \right .
	\nonum \\
	&\left . - e\bar{\psi}_\sigma \,\gamma^0 A_0\, \psi^\sigma + e \frac{v}{c}\,\bar{\psi}_\sigma\, \vec{\gamma} \cdot \vec{A}\, \psi^\sigma \right ]
\nonum \\
&+\, \int \D t\, \D^{D_\gamma} x\,\left[ - \frac{1}{4}\,F^{\mu \nu}\,F_{\mu \nu} - \frac{1}{2\xi}\left(\partial_{\mu}A^{\mu}\right)^2 \right]\, ,
\label{chap1:model-general}
\end{flalign}
where $\psi^\sigma \equiv \psi^\sigma(t,\vec x\,)$ is a four component spinor field of spin index $\sigma$ which varies from $1$ to $N_F$ ($N_F=2$ for graphene),
$v$ is the Fermi velocity, $c$ is the velocity of light which is also implicitly contained in the gauge field action through $\partial_\mu = (\frac{1}{c}\partial_t,\vec \nabla\,)$,
$\xi$ is the gauge fixing parameter and $\gamma^\mu$ is a $4\times 4$ Dirac matrix satisfying the
usual algebra: $\{ \gamma^\mu,\gamma^\nu \} = 2 g^{\mu \nu}$
where $g^{\mu \nu} = {\rm diag}(1,-1,-1,\cdots,-1)$ is the metric tensor in $D_e+1$-dimensions.
The action (\ref{chap1:model-general}) describes the coupling of a fermion field in $d_e=D_e+1$-dimensions with a $U(1)$ gauge field in $d_\gamma = D_\gamma + 1$-dimensions.
In the case of graphene, we have: $D_e=2$ and $D_\gamma = 3$, \ie, fermions in the plane and gauge field in the bulk.
Because of the running of $v$ all the way up to $c$, any complete renormalization group analysis of Dirac materials should in principle
be based on (\ref{chap1:model-general}). 
It turns out that such a task is rather involved and, presently, very few results are available, see \eg, Refs.~[\onlinecite{Gonzalez:1993uz,PhysRevLett.116.116803}].

In the literature, the overwhelming majority of works on Dirac liquids focus on the non-relativistic limit where $v/c \ra 0$ (instantaneous interactions).
This is of course, a very realistic assumption given the smallness of the ratio, \eg, $v/c \approx 1/300$ for graphene, at the experimentally accessible scales.
In this limit, there is no coupling to vector photons and Eq.~(\ref{chap1:model-general}) simplifies as (model II):
\begin{flalign}
	S &= \int \D t\, \D^{D_e} x\, \bar{\psi}_\sigma \left[ \gamma^0 \left( \I \partial_t -eA_0 \right) + \I v \vec{\gamma} \cdot \vec{\nabla}\,\right] \psi^\sigma
	\nonum \\
	&+\, \frac{1}{2}\,\int \D t\, \D^{D_\gamma} x\, \left( \vec{\nabla} A_0 \right)^2 \, ,
\label{chap1:model-inst}
\end{flalign}
where the Coulomb gauge is used. Most of the theoretical results derived on the basis of (\ref{chap1:model-inst}) are perturbative with expansions in the (bare) 
coupling constant reaching two-loop accuracy (some partial results are available at three-loop \cite{PhysRevB.89.235431}). 
Of course, given the strength of the interaction in this limit, \eg, $\al_g \approx 2.2$ for graphene, such expansions may not be reliable and a non-perturbative treatment of 
the interactions seems to be required. Such treatments are in general limited to an random phase approximation-like resummation or leading order (LO) in the $1/N$-expansion, 
see Ref.~[\onlinecite{PhysRevLett.113.105502}] for an attempt to compute next-to-leading order (NLO) corrections. 
Often, even LO results are approximate (using the so-called static approximation, neglecting Fermi velocity renormalization, etc...). So, despite the fact that 
(\ref{chap1:model-inst}) is simpler than (\ref{chap1:model-general}), calculations are difficult to carry out in a rigorous way in this limit. This often results in a rather confusing
situation where even the simplest quantities are subject to theoretical uncertainties as mentioned in the Introduction, see, \eg, 
Refs.~[\onlinecite{Kotikov:2016yrn,Teber:2014ita}] for examples and references therein.

In this paper, we will follow an alternative non-conventional route initiated in Refs.~[\onlinecite{{Teber:2012de,Kotikov:2013kcl,Kotikov:2013eha}}]. 
We will study interaction effects starting from the IR Lorentz invariant
fixed point where $v/c \ra 1$ and the interaction is fully retarded. In this limit, Eq.~(\ref{chap1:model-general}) can be written in covariant form as (model III):
\begin{flalign}
	S &= \int \D^{d_e} x\, \bar{\psi}_\sigma \I \slashed D  \psi^\sigma 
	\nonum \\
	&+ \int \D^{d_\gamma} x\,\left[ - \frac{1}{4}\,F^{\mu \nu}\,F_{\mu \nu} - \frac{1}{2\xi}\left(\partial_{\mu}A^{\mu}\right)^2 \right]\, ,
\label{chap1:rqed}
\end{flalign}
where $D_\mu = \partial_\mu + \I e A_\mu$ is the covariant derivative. As anticipated in the Introduction, we will refer to this model as
reduced QED~\cite{Gorbar:2001qt} or QED$_{d_\gamma,d_e}$ for short. For $d_e=d_\gamma =d$, Eq.~(\ref{chap1:rqed}) simply reduces to usual QED$_d$. 
The peculiar case of QED$_{4,3}$ describes graphene, and other planar Dirac liquids, at its Lorentz invariant fixed point. In this respect, model II 
corresponds to a non-relativistic reduced QED$_{d_\gamma,d_e}$  (NRQED$_{d_\gamma,d_e}$) while model III interpolates between I and II. 
 From the field theoretic point of view, the model of Eq.~(\ref{chap1:rqed}) (and similarly for the two previous ones) is characterized by an
effective free gauge-field action with fractional d'Alembertian.\footnote{The appearance of fractional d'Alembertian in Eq.~(\ref{chap1:rqed-int}) for $\veps_e>0$ 
implies that the reduced theory is non-local. In the terminology of Gracey~\cite{Gracey:2006jc} the action of the reduced gauge field seems to be 
a ``localizable non-locality'' because it can be written as a finite number of local operators as in Eq.~(\ref{chap1:rqed}).}$^,\,$\footnote{Fractional d'Alembertians 
(or Laplacians) appear in the field of fractional calculus, see Ref.~[\onlinecite{samko1993fractional}] for an extended monograph. In mathematics, 
there is a trick apparently due to Caffarelli and Silvestre~\cite{Caffarelli07} which amounts to re-write the fractional field theory in $d$-dimensional space 
as a local theory in a $d+1$-dimensional space; in our frame, this is nothing else but simply going from (\ref{chap1:rqed-int}) back to (\ref{chap1:rqed})  in the peculiar case $\veps_e=1/2$. 
See also Ref.~[\onlinecite{Rajabpour:2011qr}] for a nice account on the conformal invariance of (``localizable'') non-local field theories and 
Refs.~[\onlinecite{Limtragool:2016gnl,LaNave:2017lwf,LaNave:2017nex}] for recent references on non-local QFTs 
which make explicit use of the Caffarelli-Silvestre trick and applications to cuprates.}    
 The later can be derived from Eq.~(\ref{chap1:rqed}) by integrating out the gauge degrees of freedom
transverse to the $d_e$-dimensional manifold. Including fermions, the Lagrangian density $\mathcal{L}$ which is such that: $S =  \int \D^{d_e} x\, \mathcal{L}$, reads:
\begin{flalign}
	&\mathcal{L} = \bar{\psi}_\sigma \I \bigg( \slashed{\partial} + \I e \tilde{\slashed{A}} \bigg)  \psi^\sigma 
	- \frac{1}{4}\,\tilde{F}^{\mu \nu}\,\frac{(4\pi)^{\veps_e}}{\Gamma(1-\veps_e)\,[-\Box \,]^{\veps_e}}\,\tilde{F}_{\mu \nu} 
	\nonum \\
	&+ \frac{1}{2\tilde{\xi}}\,\tilde{A}^\mu \frac{(4\pi)^{\veps_e}\,\partial_{\mu} \partial_\nu}{\Gamma(1-\veps_e)\,[-\Box \,]^{\veps_e}}\,\tilde{A}^{\nu}\, ,
\label{chap1:rqed-int}
\end{flalign}
where we used the notation $\tilde{A}^\mu$ to emphasize the fact that it is a reduced gauge field (in $d_e$-dimensional space), 
$\veps_e = (d_\gamma-d_e)/2$ and $\tilde{\xi} = \veps_e + (1-\veps_e)\, \xi$, see Sec.~\ref{Sec:sub:set-up} for more on notations.  
Though {\it a priori} mainly of academic interest, the general motivation to consider reduced QED models is that relativistic invariance 
allows for a rigorous and systematic study of interaction effects as explained in the Introduction. We will therefore focus on a field-theoretic renormalization 
study of model III as a prerequisite to study model II and eventually model I.~\footnote{Such an approach is quite general and may also be applied to phenomena with external
fields not mentioned in the Introduction such as, \eg, magnetic catalysis, see Ref.~[\onlinecite{Miransky:2015ava,Gusynin:2013noa}] for reviews and, \eg, 
Refs.~[\onlinecite{DeTar:2016vhr,DeTar:2016dmj}] for recent works related to model II.}

\subsection{Model and conventions}
\label{Sec:sub:set-up}

\begin{figure}
	\includegraphics[scale=0.9]{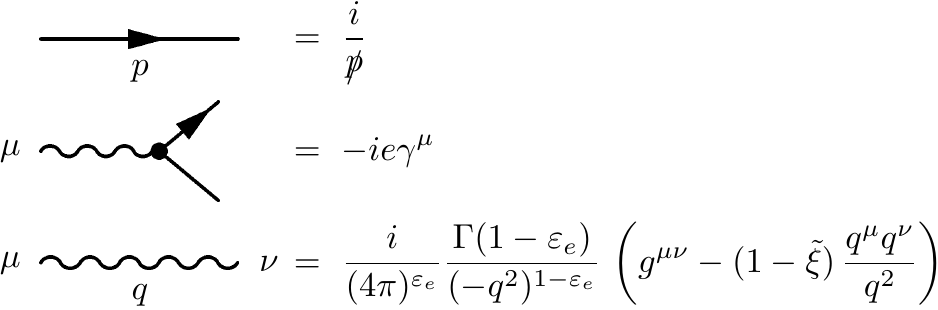}
    \caption{\label{chap3:RQED:fig:FeynmanRules}
    Feynman rules for massless reduced QED$_{d_\gamma,d_e}$ (model III).}
\end{figure}

We now proceed on presenting the model and setting up our conventions and notations, see also Refs.~[\onlinecite{Teber:2012de,Kotikov:2013eha}]. 
The Feynman rules for model III, Eq.~(\ref{chap1:rqed}), are summarized in Fig.~\ref{chap3:RQED:fig:FeynmanRules}.
The free massless fermion propagator and fermion-photon vertex are the standard ones:
\be
S_0(p) = \frac{\I}{\Sp}, \qquad \Gamma_0^\mu  = \gamma^\mu,
\label{chap2:RQED:FR:S0+Gamma0}
\ee
%
and the reduced gauge field propagator reads (see also Eq.~(\ref{chap1:rqed-int}) where fractional powers appear explicitly at the level of the action):
\be
\tilde{D}_0^{\mu \nu}(q) = \frac{\I}{(4\pi)^{\varepsilon_e}}\frac{\Gamma(1-\varepsilon_e)}{(-q^2)^{1-\varepsilon_e}}\,\left( g^{\mu \nu} - (1-\tilde{\xi})\,\frac{q^{\mu} q^{\nu}}{q^2}\right),
\label{chap3:RQED:FR:Dmunu0}
\ee
where all components of momentum as well as the indices take their values in the $d_e$-dimensional space. 
The gauge fixing parameter of the reduced gauge field, $\tilde{\xi} = 1 - \tilde{\eta}$, is
related to the gauge fixing parameter of the four-dimensional gauge field, $\xi=1-\eta$, with the help of:
\be
\tilde{\xi} = \veps_e + (1-\veps_e)\, \xi, \qquad \tilde{\eta} = (1-\veps_e)\, \eta\, .
\ee
The photon propagator, Eq.~(\ref{chap3:RQED:FR:Dmunu0}), can be separated in longitudinal and transverse parts which read:
\begin{subequations}
\label{chap2:RQED:FR:Dperp+Dpara}
\begin{flalign}
&\tilde{d}_{0 \parallel}(q^2) = \frac{\I \tilde{\xi}}{(4\pi)^{\varepsilon_e}}\frac{\Gamma(1-\varepsilon_e)}{(-q^2)^{1-\varepsilon_e}}\, ,
\label{chap2:RQED:FR:Dpara}\\
&\tilde{d}_{0 \bot}(q^2) = \frac{\I}{(4\pi)^{\varepsilon_e}}\frac{\Gamma(1-\varepsilon_e)}{(-q^2)^{1-\varepsilon_e}}\, .
\label{chap2:RQED:FR:Dperp}
\end{flalign}
\end{subequations}
In the case of QED$_{4,3}$: $\varepsilon_e=1/2$ and the reduced propagator has a square root branch-cut whereas for
QED$_{4,2}$: $\varepsilon_e=1$ and the reduced propagator is logarithmic. Notice that reduced QED$_{4,2}$ models a one-dimensional system where fermions interact via the long-range (fully retarded)
Coulomb interaction.~\cite{Gorbar:2001qt,Kaplan:2009kr}
Another case is that of QED$_{4,1}$: $\varepsilon_e=3/2$ which corresponds to a point-like particle in a four-dimensional electromagnetic environment.
In all cases the reduced QFT is non-local. 

Switching on interactions, the dressed fermion propagator and fermion-photon vertex take the form:
\begin{subequations}
\label{chap3:RQED:Dyson:S+Gamma}
\begin{flalign}
&S(p) = \frac{\I}{\Sp}\frac{1}{1-\Sigma_V(p^2)}\, ,
\label{chap3:RQED:Dyson:S} \\
&\Gamma^\mu (p,p') = \gamma^\mu + \Lambda^\mu (p,p')\, ,
\label{chap3:RQED:Dyson:Gamma}
\end{flalign}
\end{subequations}
where $\Sigma_V(p^2)$ is defined from the fermion-self-energy:
\be
\Sigma(p) = \Sp \Sigma_V(p^2)\, , 
\label{chap3:RQED:def:Sigma_V}
\ee
in a form appropriate to the massless case.
As for the photon propagator, only its transverse part is affected by interactions as in usual QED (with its precise form depending on $\veps_e$):
\begin{subequations}
\label{chap3:RQED:Dyson:D}
\begin{flalign}
&\tilde{d}_{\parallel}(q^2) = \tilde{d}_{0\,\parallel}(q^2)\, , 
\\
&\tilde{d}_{\bot}(q^2) = \tilde{d}_{0 \, \bot}(q^2) \frac{1}{1- \I q^2 \, \tilde{d}_{0\, \bot}(q^2)\, \Pi(q^2)}\, ,
\end{flalign}
\end{subequations}
where $\Pi(q^2)$ is defined from the photon self-energy:
\be
\Pi^{\mu \nu}(q) = \big(\, g^{\mu \nu} q^2 - q^\mu q^\nu\,\big)\,\Pi(q^2)\, .
\label{def:pimunu}
\ee

For arbitrary $d_e$ and $d_\gamma$ the dimensions of the fields and coupling are given by:
\begin{subequations}
\label{chap3:dims}
\begin{flalign}
&[\psi] = \frac{d_e-1}{2} = \frac{3}{2} - \varepsilon_e - \varepsilon_\gamma, \qquad [A^\mu] = \frac{d_\gamma - 2}{2} = 1 - \varepsilon_\gamma \, ,
\label{chap3:dims:psi+A}\\
&[e] = 2- \frac{d_\gamma}{2} = \varepsilon_\gamma\, ,
\label{chap3:dims:e}
\end{flalign}
\end{subequations}
where the parameters $\varepsilon_\gamma$ and $\varepsilon_e$ read:
\bea
\varepsilon_\gamma = \frac{4-d_\gamma}{2}, \qquad \varepsilon_e = \frac{d_\gamma - d_e}{2}\, .
\label{chap3:params}
\eea
Alternatively, the dimensions can be expressed as:
\be
d_\gamma = 4 -2\veps_\gamma, \qquad d_e = 4 - 2\veps_e - 2 \veps_\gamma\, .
\label{chap3:def:de+dg}
\ee
Notice that, in momentum-space, we have: 
\be
[\tilde{A}^\mu (k)] = \frac{d_e}{2} - [A^\mu] = 1 - \veps_e\, ,
\ee
and $\veps_e$ therefore appears as an anomalous dimension for the reduced gauge-field in accordance with the form of the reduced gauge-propagator,
see Fig.~\ref{chap3:RQED:fig:FeynmanRules}. Accordingly, gauge fixing is non-local with the usual factor $1/q^4$ appearing in factor of $\xi$
replaced by $1/(q^2)^{2-\veps_e}$ for arbitrary $\veps_e$. Hence, upon performing a gauge transformation: $A_\mu(x) \ra A_\mu(x) + \partial_\mu \varphi(x)$,
the correlator of the $\varphi$-field, which is proportional to the longitudinal part of the photon Green's function, also becomes anomalous.
Formally, this amounts to defining a non-local gauge transformation for the reduced gauge-field:
\be
\tilde{A}_\mu(x) \ra \tilde{A}_\mu(x) + \partial_\mu^{1+\veps_e} \tilde{\varphi}(x)\, ,
\ee
where $\partial_\mu^{1+\veps_e}$ is a fractional derivative, see, \eg, Ref.~[\onlinecite{Shirkov:1989bp}].~\footnote{It seems therefore that reduced QED 
provides a simple and concrete example of the non-local field theories studied in the recent Refs.~[\onlinecite{LaNave:2017lwf,LaNave:2017nex}] 
and invoked phenomenologically in Ref.~[\onlinecite{Limtragool:2016gnl}] without knowing the corresponding higher dimensional model.}

Despite being non-local, all reduced models with a $4$-dimensional gauge field ($d_\gamma=4$) are renormalizable as witnessed by the fact that 
the coupling constant is dimensionless in QED$_{4,d_e}$ whatever space the fermion field lives in, see Eq.~(\ref{chap3:dims:e}).
This is in agreement with the counting of ultraviolet (UV) divergences as the degree of divergence of a diagram $G$, $\om(G)$, in QED$_{d_\gamma,d_e}$ 
does not depend on the number of vertices whatever value $d_e$ takes.~\cite{Teber:2012de,Kotikov:2013eha} Moreover, 
the most superficially divergent amplitudes in QED$_{4,d_e}$s are the fermion self-energy and the fermion-gauge vertex: $\om(\Sigma_V) = 0$ and $\om(\Gamma)=0$, respectively, \ie, 
they are logarithmically divergent as in QED$_{4}$.  On the other hand, the degree of divergence of the photon self-energy is lowered in reduced QEDs;
while it is logarithmic in QED$_{4}$, $\om(\Pi)=0$, it is convergent in RQED$_{4,3}$: $\om(\Pi)=-1$, where $\Pi \equiv \Pi(q^2)$, see Eq.~(\ref{def:pimunu}). 
At this point, it is important to note that, according to 
Weinberg's theorem~\cite{Weinberg:1959nj}, a Feynman graph $G$ is absolutely convergent not only if its degree of divergence, $\om(G)$, 
is negative but also if the degrees of divergence, $\om(\gamma)$, associated to all of its subgraphs $\gamma$ are also negative. When considering multi-loop diagrams, 
one often encounters diagrams with divergent subgraphs and dealing with these subdivergences is one of the central aspect of renormalization theory.
This will be our focus in the following, with an extreme case in QED$_{4,3}$ corresponding to an overall finite graph (the photon self-energy) 
with divergent (fermion and fermion-gauge vertex) subgraphs.

We are now in a position to introduce the renormalization constants associated with a general model of QED$_{4,d_e}$:
\begin{subequations}
\begin{flalign}
&\psi = Z_{\psi}^{1/2} \psi_r, \qquad A = Z_A^{1/2} A_r\, , 
\\
&e = Z_e e_r \mu^{\veps_\gamma} = \frac{Z_\Gamma}{Z_\psi Z_A^{1/2}}\,\, e_r \mu^{\veps_\gamma}, \quad \xi =  Z_\xi \xi_r\, ,
\label{chap3:RQED:Z}
\end{flalign}
\end{subequations}
where the subscript $r$ denotes renormalized quantities and the renormalization scale, $\mu$, has been introduced in such a way that $e_r$ is dimensionless in $d_\gamma=4-2\veps_\gamma$ dimensions.
The latter is related to the corresponding parameter $\overline{\mu}$ in the modified minimal subtraction ($\overline{\text{MS}}$) scheme with the help of:
\be
\overline{\mu}^{\,2} = 4\pi e^{-\gamma_E} \mu^2\, ,
\label{chap2:muMSbar}
\ee
where $\gamma_E$ is Euler's constant. The renormalization constants also relate renormalized and bare propagators as follows:
\begin{subequations}
\label{chap2:renormalized-propagators}
\bea
S(p;\al,\xi) &=& Z_\psi(\al_r) S_r(p;\al_r,\xi_r,\mu)\, ,
\label{chap2:renormalized-S} \\
D^{\mu \nu} (q;\al,\xi) &=& Z_A(\al_r) D_r^{\mu \nu} (q;\al_r,\xi_r,\mu)\, ,
\label{chap2:renormalized-D}\\
\Gamma^\mu(p,p'; \al,\xi) &=& Z_\Gamma^{-1} (\al_r) \Gamma_r^\mu(p,p'; \al_r,\xi_r, \mu)  \, ,
\label{chap2:renormalized-Gamma}
\eea
\end{subequations}
where the bare propagators do not depend on $\mu$. In the $\text{MS}$ scheme, these constants take the simple form:
\be
Z_x(\al_r,\xi_r) = 1 + \delta Z_x (\al_r,\xi_r) = 1 + \sum_{l=1}^\infty \sum_{j=1}^l Z_x^{(l,j)}(\xi_r)\,\frac{\al_r^l}{\veps_\gamma^j}\, ,
\label{chap2:model:QED_d:Z-exp}
\ee
where $x \in \{ \psi,A,e,\xi,\Gamma \}$, $\al_r = e_r^2/(4\pi)$  and $l$ runs over the number of loops at which UV singularities are subtracted. 
In the $\text{MS}$ scheme the $Z_x$ do not depend on momentum or mass; furthermore,
the dependence on $\mu$ is only through $\al_r$ and/or $\xi_r$. So the $Z_x$ depend only on $\al_r(\mu)$, $\veps_\gamma$ and eventually $\xi_r(\mu)$. 

From the renormalization constants, it is possible to compute the $\beta$-function:
\be
\beta(\al_r) = \mu \, \frac{\partial \al_r}{\partial \mu}\bigg|_B \qquad (Z_\al = Z_e^2)\, ,
\label{chap2:model:QED_d:beta}
\ee
where the subscript $B$ indicates that bare parameters, which do not depend on $\mu$, are fixed. Explicitly, it reads:
\be
\beta(\al_r) = -2\veps_\gamma \al_r + \sum_{l=0}^\infty \beta_l \al_r^{l+2}, \qquad \beta_l = 2(l+1)\,Z_\al^{(l+1,1)}\, ,
\label{chap2:model:QED_d:beta3}
\ee
where the coefficients $\beta_l$ are completely determined by the simple $1/\veps_\gamma$ poles in $Z_\al$. Similarly, one may compute the field anomalous dimensions
which are defined as:
\be
\gamma_x(\al_r,\xi_r) = - \mu \,\frac{\D \log Z_x(\al_r,\xi_r)}{\D \mu}\bigg|_B \qquad (x \in \{ \psi,A \})\, .
\label{chap2:model:QED_d:anomalousdims}
\ee
The radiatively generated photon anomalous dimension is gauge invariant and reads:
\be
\gamma_A(\al_r) = \sum_{l=0}^\infty \gamma_{A,l} \al_r^{l+1}, \quad \gamma_{A,l}= 2(l+1)\,Z_A^{(l+1,1)}\, .
\label{chap2:model:QED_d:anomalousdims-A}
\ee
In the case of the fermion anomalous dimension, we have:
\be
\gamma_\psi(\al_r,\xi_r) = \sum_{l=0}^\infty \gamma_{\psi,l}(\xi_r) \al_r^{l+1}, \quad \gamma_{\psi,l}(\xi_r)= 2(l+1)\,Z_\psi^{(l+1,1)}(\xi_r)\, .
\label{chap2:model:QED_d:anomalousdims-psi}
\ee

As in usual QEDs, the renormalized constant are not all independent. The gauge non-invariant gauge-fixing term is not renormalized, hence: $Z_\xi = Z_A$.
Moreover, the Ward identity:
\be
Z_\psi = Z_\Gamma\, ,
\label{chap3:WI:Zpsi}
\ee
holds~\cite{Teber:2012de} for arbitrary $d_e$ implying that $Z_e = Z_A^{-1/2}$. Finally, the free gauge-field action is non-local in the reduced case
and hence the gauge-field is not renormalized:~\cite{Vasil'evbook} $Z_A = 1$, which implies that $Z_e=1$. As a consequence, there is no radiatively generated photon anomalous dimension and
the $\beta$-function is zero which implies that the coupling remains marginal to
all orders in perturbation theory:
\be
\beta(\al_r) = 0 \qquad d_e <4\, ,
\ee
a fact reminiscent of the ($1+1$)-dimensional Tomonaga-Luttinger model.~\cite{PhysRevLett.67.3852,PhysRevB.47.16107}
 Assuming that the coupling constant is weak enough that no dynamical mass is generated,~\cite{Gorbar:2001qt,Kotikov:2016yrn} reduced QED is therefore conformally invariant.

As anticipated in the Introduction, in order to compute renormalization constants and renormalized correlators, we will use the
BPHZ prescription:~\cite{Bogoliubov:1957gp,Hepp:1966eg,Zimmermann1969} 
\begin{subequations}
\label{chap2:def:forest}
\bea
\mathcal{R}\,G &=& (1 - \mathcal{K})\,\mathcal{R}'\,G\, ,
\label{chap2:def:R}\\
\mathcal{R}'\,G &=& G + \sum_{\bar{\Gamma}_d \not= \emptyset} \prod_{\gamma \in \bar{\Gamma}_d} \bigg( - \mathcal{K} \mathcal{R}' \gamma \bigg) \, \star \, G/\bar{\Gamma}_d\, ,
\label{chap2:def:R'}
\eea
\end{subequations}
where  $\mathcal{R}\,G$ corresponds to the finite (renormalized) graph $G$
with all divergences (both subdivergences and overall divergence) subtracted. In Eqs.~(\ref{chap2:def:forest}), 
$\mathcal{R}'$ is the so-called incomplete $R$-operation because it subtracts only the subdivergences, 
$\bar{\Gamma}_d$ is the set of all subdivergent graphs which are disjoint (nested ones are not allowed) and the operator $\mathcal{K}$ is defined as: 
\be
\mathcal{K}~\left( \sum_{n=-\infty}^{+\infty} \frac{c_n}{\veps_\gamma^n} \right) = \sum_{n=1}^{+\infty} \frac{c_n}{\veps_\gamma^n}\, .
\label{chap2:def:Sing}
\ee
Moreover, the notation $G/\bar{\Gamma}$ means that the subdiagrams contained in $G$ are shrunk to a vertex
and the $\star$ operation amounts to substitute the counterterm in the integrand of the shrunk diagram. In the case of (at most) logarithmic graphs, 
the $\star$ operator reduces to simple multiplication and we may identify the counter-term with the renormalization constant, \ie, $Z(\gamma) = \mathcal{K} \mathcal{R}'\gamma$.

\section{Reduced QED at one loop}
\label{Sec:One-Loop}

\begin{figure}
    \includegraphics{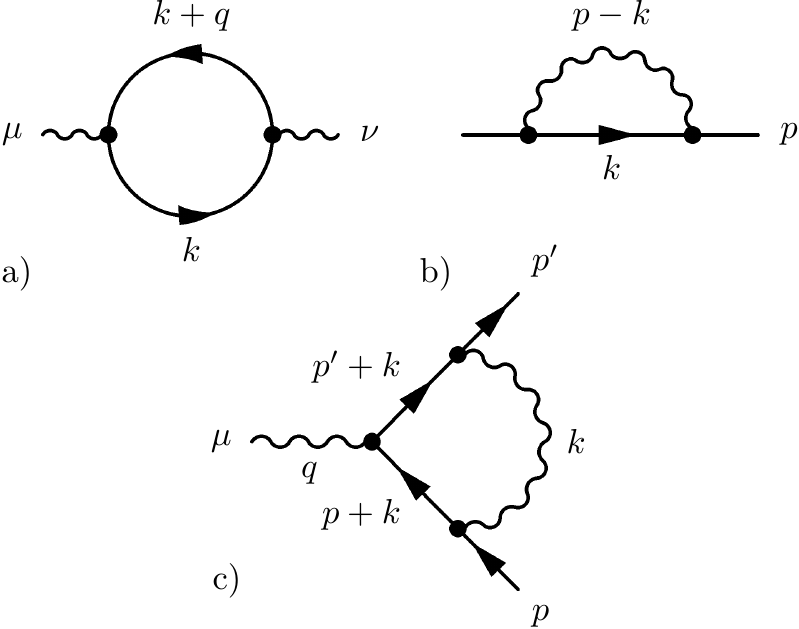}
  \caption{\label{chap2:fig:one-loop}
  One-loop diagrams: a) gauge field self-energy, b) fermion self-energy and c) fermion-gauge field vertex.}
\end{figure}

We now proceed on reviewing the one-loop structure of the model.~\cite{Teber:2012de}
The one-loop fermion self-energy, polarization operators and fermion-gauge field vertex, see Fig.~\ref{chap2:fig:one-loop}, are defined as:
\begin{subequations}
\label{chap3:def:Pi1+Sigma1+Lambda1}
\begin{flalign}
&\I \Pi_1^{\mu \nu}(q) =
-\int [\D^{d_e} k] \Tr \bigg[ (-\I e \gamma^\mu)\,S_0(k)\,(-\I e \gamma^\nu)\,S_0(k+q) \bigg]\, ,
\label{chap3:def:Pi1}\\
&-\I \Sigma_1(p) = \int [\D^{d_e} k] (-\I e \gamma^\mu)\,\tilde{D}_{0,\mu \nu}(p-k)\,S_0(k)\,(-\I e \gamma^\nu)\, ,
\label{chap3:def:Sigma1} \\
&-\I e \Lambda_1^{\mu}(p,p') = \int [\D^{d_e} k] \tilde{D}_0^{\al \beta}(p-k)\,(-\I e \gamma_\al)\,S_0(k)\times
\nonum \\
&\qquad \times\,(-\I e \gamma^\mu)\,S_0(k+q)\,(-\I e \gamma_\beta)\, .
\label{chap3:def:Lambda1}
\end{flalign}
\end{subequations}
The following parametrizations are useful:
\be
\Pi(q^2) = \frac{- \Pi^\mu_\mu(q)}{(d_e-1)\,(-q^2)}, \quad \Sigma_V(p^2) = \frac{-1}{4N_F\,(-p^2)}\,\Tr[ \Sp \Sigma(p)]\, .
\ee
All calculations done, see App.~\ref{App:MI} for the master integrals, the results read:
\begin{subequations}
\label{chap3:gen-expr:Pi1+Sigma1+Lambda1}
\begin{flalign}
&\Pi_1(q^2;\bar{\al}) = - 4 N_F \bar{\al} \left( \frac{4\pi}{-q^2} \right)^{\veps_e} \,\left( \frac{\overline{\mu}^{\,2}}{-q^2} \right)^{\veps_\gamma}\, \times
	\nonum \\
	&\frac{d_e-2}{2(d_e-1)}\,e^{\gamma_E \veps_\gamma} G(d_e,1,1)\, ,
\label{chap3:res:Pi1} \\
&\Sigma_{1V}(p^2) = \bar{\al} \,\left( \frac{\overline{\mu}^{\,2}}{-p^2} \right)^{\veps_\gamma}\, \Gamma(1-\veps_e)\,\frac{d_e-2}{2}\,\times
	\nonum \\
	&\bigg( \frac{\veps_e}{d_e-2+\veps_e} - \xi\,\bigg)\,e^{\gamma_E \veps_\gamma} G(d_e,1,1-\veps_e)\, ,
\label{chap3:res:Sigma1} \\
&\Lambda_1^{\mu}(p=p'=0) = \bar{\al}\,\gamma^\mu\,\left( \frac{\overline{\mu}^{\,2}}{m^2} \right)^{\veps_\gamma}\,\times
	\nonum \\
	&\bigg( \frac{(d_e-2)^2}{d_e(1-\veps_e)} - (1 - \xi) \bigg)\,e^{\gamma_E \veps_\gamma}\,\frac{\Gamma(1+\veps_\gamma)}{\veps_\gamma} \, ,
\label{chap3:res:Lambda1}
\end{flalign}
\end{subequations}
where $\bar{\al}=\al/(4\pi)$, we have used the fact that $(1-\tilde{\xi}) = (1-\veps_e)\,(1-\xi)$ and the vertex has been computed at 
$p=p'=0$ with a small mass regulating an IR singularity. For later purposes, let's recall that the photon propagator (internal line in $\Sigma_1$) has
a longitudinal and a transverse part, see Eq.~(\ref{chap2:RQED:FR:Dperp+Dpara}). Then, Eq.~(\ref{chap3:res:Sigma1}) shows that a similar decomposition holds for the one-loop fermion self-energy:
\begin{subequations}
\label{chap3:Sigma1:LT}
\begin{flalign}
\Sigma_{1V}(p^2) &= \Sigma_{1V}^{(\parallel)}(p^2) + \Sigma_{1V}^{(\perp)}(p^2) \, ,
\label{chap3:Sigma1:L+T}\\
\Sigma_{1V}^{(\parallel)}(p^2) &= -\xi \,\bar{\al} \,\left( \frac{\overline{\mu}^{\,2}}{-p^2} \right)^{\veps_\gamma}\, \Gamma(1-\veps_e)\,\times
	\nonum \\
	&\times\,\frac{d_e-2}{2}\, e^{\gamma_E \veps_\gamma} G(d_e,1,1-\veps_e)\, ,
\label{chap3:Sigma1:L}\\
\Sigma_{1V}^{(\perp)}(p^2) &= \veps_e \, \bar{\al} \,\left( \frac{\overline{\mu}^{\,2}}{-p^2} \right)^{\veps_\gamma}\, \Gamma(1-\veps_e)\,\times
	\nonum \\
	&\times\,\frac{d_e-2}{2(d_e-2+\veps_e)}\,e^{\gamma_E \veps_\gamma} G(d_e,1,1-\veps_e)\, ,
\label{chap3:Sigma1:T}
\end{flalign}
\end{subequations}
where the transverse part is non-zero only in the reduced case.

Focusing on QED$_{4,d_e}$, the singular part of these self-energies allows to extract the one-loop counter-terms
which read:
\begin{subequations}
\label{chap2:RPT:counterterms}
\begin{flalign}
	&\delta Z_{1A}(\bar{\al}_r) = \mathcal{K}\bigg[ \Pi_{1}(q^2;\bar{\al}_r) \bigg] 
	\nonum \\
	&= \mathcal{K}\bigg[~~
      \parbox{15mm}{
      \begin{fmfgraph*}(15,15)
      \fmfleft{in}
      \fmfright{out}
      \fmf{boson}{in,ve}
      \fmf{plain,right,tension=0.2}{ve,vw}
      \fmf{plain,right,tension=0.2}{vw,ve}
      \fmf{boson}{vw,out}
      \fmfdot{ve,vw}
    \end{fmfgraph*}
}~~
	\bigg] = 0 \quad (d_e < 4)\, ,
\label{chap2:RPT:counterterms:A}\\
	&\delta Z_{1\psi}(\bar{\al}_r,\xi_r) = \mathcal{K}\bigg[ \Sigma_{1V}(p^2;\bar{\al}_r,\xi_r) \bigg] 
	\nonum \\
	&=\mathcal{K}\bigg[~~
      \parbox{15mm}{
    \begin{fmfgraph*}(15,15)
      \fmfleft{in}
      \fmfright{out}
      \fmf{plain}{in,vi}
      \fmf{plain,tension=0.2}{vi,vo}
      \fmf{boson,left,tension=0.2}{vi,vo}
      \fmf{plain}{vo,out}
      \fmfdot{vi,vo}
    \end{fmfgraph*}
}~~
\bigg] = \bar{\al}_r \,\bigg( \frac{\veps_e}{2-\veps_e} - \xi_r \bigg)\,\frac{1}{\veps_\gamma}\, ,
\label{chap2:RPT:counterterms:S}\\
	&\delta Z_{1\Gamma}(\bar{\al}_r,\xi_r) =  - \mathcal{K}\bigg[\Lambda_{1}^\mu(p,p';\bar{\al}_r,\xi_r)/ \gamma^\mu \bigg]
	\nonum \\
	&= - \mathcal{K}\bigg[~~
      \parbox{15mm}{
    \begin{fmfgraph*}(15,15)
      \fmfleft{in}
      \fmfright{e1,e2}
      \fmf{boson}{in,vi}
      \fmf{plain}{e1,v1}
      \fmf{plain,tension=0.7}{v1,vi}
      \fmf{plain,tension=0.7}{vi,v2}
      \fmf{plain}{v2,e2}
      \fmffreeze
      \fmf{boson,right,tension=0.7}{v1,v2}
      \fmfdot{vi,v1,v2}
    \end{fmfgraph*}
}~~
\bigg] = - \bar{\al}_r \,\bigg( \frac{\veps_e}{2-\veps_e} - \xi_r \bigg)\,\frac{1}{\veps_\gamma}\, ,
\label{chap2:RPT:counterterms:G}
\end{flalign}
\end{subequations}
where the Lorentz structure of the graphs displayed in the brackets has been projected out. As anticipated in the last section, the Ward identity Eq.~(\ref{chap3:WI:Zpsi}) is satisfied 
for all $d_e$. In graphical form, the latter reads at one-loop:
\be
\mathcal{K}\bigg[ ~
      \parbox{15mm}{
    \begin{fmfgraph*}(15,15)
      \fmfleft{in}
      \fmfright{e1,e2}
      \fmf{boson}{in,vi}
      \fmf{plain}{e1,v1}
      \fmf{plain,tension=0.7}{v1,vi}
      \fmf{plain,tension=0.7}{vi,v2}
      \fmf{plain}{v2,e2}
      \fmffreeze
      \fmf{boson,right,tension=0.7}{v1,v2}
      \fmfdot{vi,v1,v2}
    \end{fmfgraph*}
} ~ \bigg] \quad = \quad - \quad \mathcal{K}\bigg[ ~
      \parbox{15mm}{
    \begin{fmfgraph*}(15,15)
      \fmfleft{in}
      \fmfright{out}
      \fmf{plain}{in,vi}
      \fmf{plain,tension=0.2}{vi,vo}
      \fmf{boson,left,tension=0.2}{vi,vo}
      \fmf{plain}{vo,out}
      \fmfdot{vi,vo}
    \end{fmfgraph*}
} ~ \bigg]\, .
\label{chap2:ward:graphical}
\ee

Explicitly, in the specific case of reduced QED$_{4,3}$ ($\veps_e=1/2$ and $\veps_\gamma \ra 0$) the $\veps_\gamma$-expansion of Eqs.~(\ref{chap3:res:Pi1}) and (\ref{chap3:res:Sigma1}) reads:
\begin{subequations}
\label{chap3:RQED4,3:Pi1+Sigma1}
\begin{flalign}
	&\Pi_1(q^2;\bar{\al}) = - N_F\,\bar{\al}\, \frac{2\pi^2}{\sqrt{-q^2}}\,\times
	\nonum \\
	&\times\,\left [ 1 - (L_q - \log 4 + 1) \,\veps_\gamma +  \Ord(\veps_\gamma^2) \right]\, ,
\label{chap3:RQED4,3:Pi1}\\
	&\Sigma_{1V}(p^2;\bar{\al},\xi) = \bar{\al} \left [ \frac{1-3\xi}{3\veps_\gamma} - \frac{1-3\xi}{3}\,\tilde{L}_p - 2\xi + \frac{10}{9}
\right . 
\nonum \\
	&\left . + \left( \frac{1-3\xi}{6}\, ( \tilde{L}_p^2 - 7 \zeta_2 ) + 2\,\left( \xi - \frac{5}{9} \right)\,\tilde{L}_p - 8 \xi +\frac{112}{27} \right)\,\veps_\gamma \right .
	\nonum \\
	&\left . +  \Ord(\veps_\gamma^2) \right]\, ,
\label{chap3:RQED4,3:Sigma1}
\end{flalign}
\end{subequations}
where $\tilde{L}_x = L_x + \log 4$. As a trivial application of the BPHZ prescription, by combining Eqs.~(\ref{chap2:def:forest}) and
(\ref{chap3:RQED4,3:Pi1+Sigma1}), the renormalized one-loop self-energies read:
\begin{subequations}
\label{chap3:RQED4,3:Pi1r+Sigma1r}
\begin{flalign}
	&\Pi_{1r}(q^2) = - \frac{N_F e^2}{8\,\sqrt{-q^2}}\, , 
	\label{chap3:RQED4,3:Pi1r}\\
	&\Sigma_{1Vr}(p^2) = -\bar{\al}_r \left( \frac{1-3\xi_r}{3}\,\tilde{L}_p + 2\xi_r - \frac{10}{9} \right)\, .
	\label{chap3:RQED4,3:Sigma1r}
\end{flalign}
\end{subequations}

\section{Two-loop polarization operator}
\label{Sec:Two-Loop-Pi}

\begin{figure}
    \includegraphics{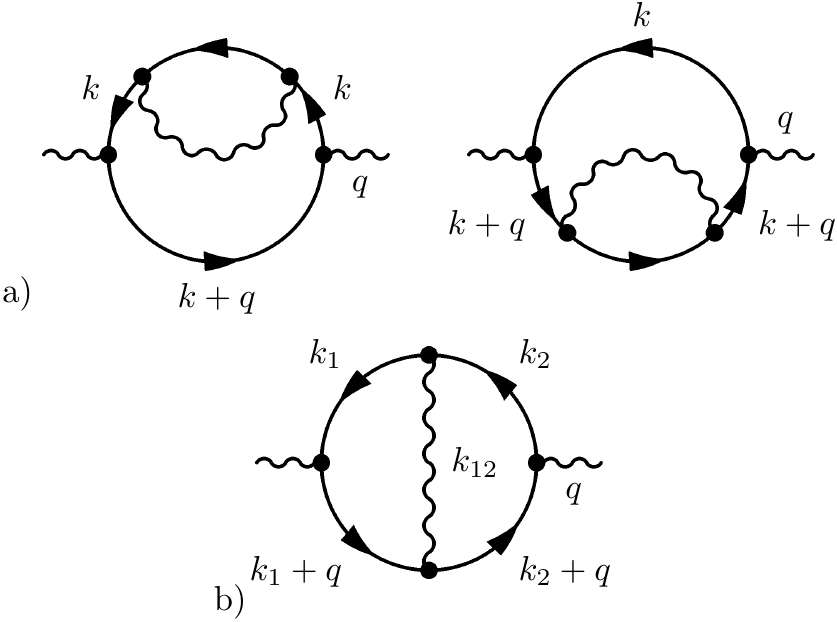}
  \caption{\label{chap3:fig:two-loop:Pi}
  Two-loop photon self-energy diagrams ($k_{12} = k_1-k_2$).}
\end{figure}

We now go on to two-loop order and first focus on the polarization operator. The total two-loop photon self-energy can be decomposed as follows:
\be
\Pi_2^{\mu \nu}(q) = 2\Pi_{2a}^{\mu \nu}(q) + \Pi_{2b}^{\mu \nu}(q)\, ,
\label{chap3:Pi2=2Pi2a+Pi2b}
\ee
where the diagrams are displayed on Fig.~\ref{chap3:fig:two-loop:Pi}. The latter are defined as:
\begin{subequations}
\label{chap3:def:Pi2}
\begin{flalign}
        &\I \Pi_{2a}^{\mu \nu}(q) = - \int [\D^{d_e} k] \Tr \bigg[ (-\I e \gamma^\mu)\,S_0(k+q)\times
        \nonum \\
        &\times\,(-\I e \gamma^\nu)\,S_0(k)\,(-\I \Sk \Sigma_{1V}(k))\,S_0(k) \bigg]\, ,
        \label{chap3:def:Pi2a}\\
        &\I \Pi_{2b}^{\mu \nu}(q) = - \int [\D^{d_e}k]\,\Tr \left[ (-\I e\gamma^\nu)\, S_0(k+q)\,\times \right .
        \nonum \\
        &\left . \times  (-\I e \Lambda_1^\mu(k,q))\, S_0(k) \right]\, ,
\label{chap3:def:Pi2b}
\end{flalign}
\end{subequations}
where the one-loop fermion self-energy and fermion-photon vertex were defined in Eqs.~(\ref{chap3:def:Pi1+Sigma1+Lambda1}).
Because $\Pi^{\mu \nu}(q)$ is gauge independent, all calculation can be carried out in a specific gauge.
In the following we shall work in the Feynman gauge, $\xi=1$.

All calculations done, the general expression for the 2-loop photon self-energy diagrams of QED$_{d_\gamma, d_e}$ read:
\begin{subequations}
\label{chap3:res:Pi2}
\begin{flalign}
        &\Pi_{2a}(q^2) = 4N_F\, \bar{\al}^2 \,\left( \frac{4\pi}{-q^2}\right)^{\veps_e}\, \left( \frac{\overline{\mu}^{\,2}}{-q^2} \right)^{2\veps_\gamma}\,\Gamma(1-\veps_e)\,\times 
 	\nonum \\
	&\times \, \frac{(d_e-2)^4}{2\,(d_e-1)\,(d_e-2+\veps_e)(d_\gamma-4)}\,\times
	\nonum \\
	&\times\,e^{2\gamma_E \veps_\gamma} G(d_e,1,1-\veps_e)G(d_e,1,\veps_\gamma)\, .
	\label{chap3:res:Pi2a} \\
	&\Pi_{2b}(q^2) = -4N_F\, \bar{\al}^2 \,\left( \frac{4\pi}{-q^2}\right)^{\veps_e}\, \left( \frac{\overline{\mu}^{\,2}}{-q^2} \right)^{2\veps_\gamma}\,\Gamma(1-\veps_e)\,\times
	\nonum \\
	&\frac{d_e-2}{2(d_e-1)}\,e^{2\gamma_E \veps_\gamma}\, \bigg [ \,2\, G(d_e,1,1-\veps_e)G(d_e,1,\veps_\gamma)\,\times \bigg .
	\nonum \\
	&\times\, \left(d_e-4 + \frac{2(d_e-2)^3}{d_e\,(d_\gamma-4)} - \frac{4(d_e-2)}{d_e+d_\gamma-6} - \frac{4(d_e-2)^2}{(d_e + d_\gamma-4)^2} + \right .
	\nonum \\
	&\left . + \frac{(d_e-2)(d_e^2-8)}{d_e(d_e + d_\gamma-4)} \right) - G(d_e,1,1,1,1,1-\veps_e)\,\times
	\nonum \\
	&\left . \times\,\left(d_e - 4 + \frac{4(d_e-2)}{d_e+d_\gamma-6} - \frac{d_e(d_e-2)}{d_e+d_\gamma-4} \right) \right ]\, ,
\label{chap3:res:Pi2c}
\end{flalign}
\end{subequations}
where the two-loop master integral $G(d_e,1,1,1,1,\al)$ with index $\al=1-\veps_e$ appears, see App.~\ref{App:MI}. In order to compute the renormalized self-energies, 
we apply the BPHZ prescription Eq.~(\ref{chap2:def:forest}). Graphically, the renormalization constants associated with each 2-loop diagram read:
\begin{subequations}
\label{chap3:def:ZA-forest}
\begin{flalign}
	&2\,\delta Z_{2a\,A}(\bar{\al}_r) = 2\,\mathcal{K}\, \bigg[~
\parbox{15mm}{
    \begin{fmfgraph*}(15,15)
      \fmfleft{i}
      \fmfright{o}
      \fmf{photon}{i,v1}
      \fmf{photon}{v2,o}
      \fmf{phantom,right,tension=0.1,tag=1}{v1,v2}
      \fmf{phantom,right,tension=0.1,tag=2}{v2,v1}
      \fmf{phantom,tension=0.1,tag=3}{v1,v2}
      \fmfdot{v1,v2}
      \fmfposition
      \fmfipath{p[]}
      \fmfiset{p1}{vpath1(__v1,__v2)}
      \fmfiset{p2}{vpath2(__v2,__v1)}
      \fmfiset{p3}{vpath3(__v1,__v2)}
      \fmfi{plain}{subpath (0,length(p1)) of p1}
      \fmfi{plain}{subpath (0,length(p2)/4) of p2}
      \fmfi{plain}{subpath (length(p2)/4,3length(p2)/4) of p2}
      \fmfi{plain}{subpath (3length(p2)/4,length(p2)) of p2}
      \fmfi{photon}{point length(p2)/4 of p2 .. point length(p3)/2 of p3 .. point 3length(p2)/4 of p2}
      \def\vert#1{%
        \fmfiv{decor.shape=circle,decor.filled=full,decor.size=2thick}{#1}}
      \vert{point length(p2)/4 of p2}
      \vert{point 3length(p2)/4 of p2}
    \end{fmfgraph*}
} ~\bigg] - 
\nonum \\
	&\quad-
2\,\mathcal{K}\, \bigg[ ~\mathcal{K}\, \bigg[~
   \parbox{15mm}{
    \begin{fmfgraph*}(15,15)
      \fmfleft{in}
      \fmfright{out}
      \fmf{plain}{in,ve}
      \fmf{plain,right,tension=0.2}{ve,vw}
      \fmf{boson,right,tension=0.2}{vw,ve}
      \fmf{plain}{vw,out}
      \fmfdot{ve,vw}
    \end{fmfgraph*}
}~ \bigg]~\star~
\parbox{15mm}{
    \begin{fmfgraph*}(15,15)
      \fmfleft{in}
      \fmfright{out}
      \fmf{boson}{in,ve}
      \fmf{plain,right,tension=0.2}{ve,vw}
      \fmf{plain,right,tension=0.2}{vw,ve}
      \fmf{boson}{vw,out}
      \fmfdot{ve,vw}
    \end{fmfgraph*}
} ~ \bigg]\, ,
\label{chap3:deltaZ2Aa}\\
	&\delta Z_{2b\,A}(\bar{\al}_r) = \mathcal{K}\, \bigg[~
\parbox{15mm}{
    \begin{fmfgraph*}(15,15)
      \fmfleft{i}
      \fmfright{o}
      \fmf{photon}{i,v1}
      \fmf{photon}{v2,o}
      \fmf{phantom,right,tension=0.1,tag=1}{v1,v2}
      \fmf{phantom,right,tension=0.1,tag=2}{v2,v1}
      \fmf{phantom,tension=0.1,tag=3}{v1,v2}
      \fmfdot{v1,v2}
      \fmfposition
      \fmfipath{p[]}
      \fmfiset{p1}{vpath1(__v1,__v2)}
      \fmfiset{p2}{vpath2(__v2,__v1)}
      \fmfi{plain}{subpath (0,length(p1)/2) of p1}
      \fmfi{plain}{subpath (length(p1)/2,length(p1)) of p1}
      \fmfi{plain}{subpath (0,length(p2)/2) of p2}
      \fmfi{plain}{subpath (length(p2)/2,length(p2)) of p2}
      \fmfi{photon}{point length(p1)/2 of p1 -- point length(p2)/2 of p2}
      \def\vert#1{%
        \fmfiv{decor.shape=circle,decor.filled=full,decor.size=2thick}{#1}}
      \vert{point length(p1)/2 of p1}
      \vert{point length(p2)/2 of p2}
    \end{fmfgraph*}
} ~\bigg] - 
\nonum \\
&\quad-2\,\mathcal{K}\, \bigg[~\mathcal{K}\, \bigg[~
   \parbox{15mm}{
    \begin{fmfgraph*}(15,15)
      \fmfleft{in}
      \fmfright{e1,e2}
      \fmf{boson}{in,vi}
      \fmf{plain}{e1,v1}
      \fmf{plain,tension=0.7}{v1,vi}
      \fmf{plain,tension=0.7}{vi,v2}
      \fmf{plain}{v2,e2}
      \fmffreeze
      \fmf{boson,right,tension=0.7}{v1,v2}
      \fmfdot{vi,v1,v2}
    \end{fmfgraph*}
}~ \bigg]~\star~
\parbox{15mm}{
    \begin{fmfgraph*}(15,15)
      \fmfleft{in}
      \fmfright{out}
      \fmf{boson}{in,ve}
      \fmf{plain,right,tension=0.2}{ve,vw}
      \fmf{plain,right,tension=0.2}{vw,ve}
      \fmf{boson}{vw,out}
      \fmfdot{ve,vw}
    \end{fmfgraph*}
} ~\bigg] \, ,
\label{chap3:deltaZ2Ac}
\end{flalign}
\end{subequations}
where, as in Eqs.~(\ref{chap2:RPT:counterterms:G}), it is understood that the Lorentz structure of the diagrams in argument of 
$\mathcal{K}$ has been projected out. Because all graphs are at most logarithmic, the $\star$ operation reduces to a simple multiplication 
and will be omitted in the following. Upon computing the total renormalization constant, we see that the last terms in Eqs.~(\ref{chap3:deltaZ2Aa}) 
and (\ref{chap3:deltaZ2Ac}) cancel eachother thanks to the Ward identity (\ref{chap2:ward:graphical}). The total two-loop renormalization constant therefore reduces to:
\bea
\delta Z_{2\,A}(\bar{\al}_r) = 2\,\mathcal{K}\, \bigg[~
\parbox{15mm}{
    \begin{fmfgraph*}(15,15)
      \fmfleft{i}
      \fmfright{o}
      \fmf{photon}{i,v1}
      \fmf{photon}{v2,o}
      \fmf{phantom,right,tension=0.1,tag=1}{v1,v2}
      \fmf{phantom,right,tension=0.1,tag=2}{v2,v1}
      \fmf{phantom,tension=0.1,tag=3}{v1,v2}
      \fmfdot{v1,v2}
      \fmfposition
      \fmfipath{p[]}
      \fmfiset{p1}{vpath1(__v1,__v2)}
      \fmfiset{p2}{vpath2(__v2,__v1)}
      \fmfiset{p3}{vpath3(__v1,__v2)}
      \fmfi{plain}{subpath (0,length(p1)) of p1}
      \fmfi{plain}{subpath (0,length(p2)/4) of p2}
      \fmfi{plain}{subpath (length(p2)/4,3length(p2)/4) of p2}
      \fmfi{plain}{subpath (3length(p2)/4,length(p2)) of p2}
      \fmfi{photon}{point length(p2)/4 of p2 .. point length(p3)/2 of p3 .. point 3length(p2)/4 of p2}
      \def\vert#1{%
        \fmfiv{decor.shape=circle,decor.filled=full,decor.size=2thick}{#1}}
      \vert{point length(p2)/4 of p2}
      \vert{point 3length(p2)/4 of p2}
    \end{fmfgraph*}
} ~\bigg] + \mathcal{K}\, \bigg[~
\parbox{15mm}{
    \begin{fmfgraph*}(15,15)
      \fmfleft{i}
      \fmfright{o}
      \fmf{photon}{i,v1}
      \fmf{photon}{v2,o}
      \fmf{phantom,right,tension=0.1,tag=1}{v1,v2}
      \fmf{phantom,right,tension=0.1,tag=2}{v2,v1}
      \fmf{phantom,tension=0.1,tag=3}{v1,v2}
      \fmfdot{v1,v2}
      \fmfposition
      \fmfipath{p[]}
      \fmfiset{p1}{vpath1(__v1,__v2)}
      \fmfiset{p2}{vpath2(__v2,__v1)}
      \fmfi{plain}{subpath (0,length(p1)/2) of p1}
      \fmfi{plain}{subpath (length(p1)/2,length(p1)) of p1}
      \fmfi{plain}{subpath (0,length(p2)/2) of p2}
      \fmfi{plain}{subpath (length(p2)/2,length(p2)) of p2}
      \fmfi{photon}{point length(p1)/2 of p1 -- point length(p2)/2 of p2}
      \def\vert#1{%
        \fmfiv{decor.shape=circle,decor.filled=full,decor.size=2thick}{#1}}
      \vert{point length(p1)/2 of p1}
      \vert{point length(p2)/2 of p2}
    \end{fmfgraph*}
} ~\bigg] \, ,
\label{chap3:deltaZ2A}
\eea
where $\delta Z_{2\,A} = 2 \,\delta Z_{2a\,A} + \delta Z_{2b\,A}$. Similarly to the case of usual QED,~\cite{itzykson2012quantum}
this simplification implies that, even though the individual diagrams have subdivergent graphs, 
the contribution of the latter cancel each other thanks to the Ward identity and their subtraction therefore does not affect the final result.~\footnote{This is the essential difference with respect 
to the non-relativistic case where the contributions of subdivergent graphs only partially cancel each-other due to non-standard Ward identities and their subtraction
therefore affects the final result, see Ref.~[\onlinecite{Teber:2018qcn}].}

We now apply the above formulas to the case of QED$_{4,3}$ ($\veps_e=1/2$ and $\veps_\gamma \ra 0$).
From Eqs.~(\ref{chap3:res:Pi2a}) and (\ref{chap3:res:Pi2c}), this leads to:
\begin{subequations}
\label{chap3:res2l:RQED4,3:Pi2a+Pi2c}
\begin{flalign}
&\Pi_{2a}(q^2) =  \frac{N_F\,\al^2}{\sqrt{-q^2}}\,\left [ \frac{1}{12\veps_\gamma} - \frac{L_q}{6} + \frac{1}{9} + \Ord(\veps_\gamma) \right]\, ,
\label{chap3:res2l:RQED4,3:Pi2a}\\
&\Pi_{2b}(q^2) =  \frac{N_F\,\al^2}{\sqrt{-q^2}}\,\left [ -\frac{1}{6\veps_\gamma} + \frac{L_q}{3} + \frac{\pi^2}{4} - \frac{25}{9} + \Ord(\veps_\gamma) \right]\, .
\label{chap3:res2l:RQED4,3:Pi2c}
\end{flalign}
\end{subequations}
While the individual contributions are divergent (with only simple poles arising from divergent subgraphs, see below), the total two-loop polarization operator is finite 
($\delta Z_{2\,A}(\al_r) = 0$) and reads:
\be
\Pi_{2}(q^2) = \Pi_{2r}(q^2) = -N_F\,\frac{\al_r^2}{\sqrt{-q^2}}\,\frac{92-9\pi^2}{36} \, .
\label{chap3:res2l:RQED4,3:Pi2}
\ee

These results can also be recovered from the computation of individual counterterms in the Feynman gauge ($\xi=1$):
\begin{subequations}
\label{chap3:res2l:RQED4,3:CT:Pi2}
\begin{flalign}
	&\delta Z_{2a\,A}(\al_r) = \mathcal{K} \mathcal{R}' \bigg[\Pi_{2a}(q^2) \bigg] 
	\nonum \\
	&= \mathcal{K} \bigg[ \Pi_{2a}(q^2) \bigg] - \mathcal{K} \bigg[ \mathcal{K} \big[ \Sigma_{1V}(\al_r) \big]\,\Pi_{1}(q^2) \bigg] 
	\nonum \\
	&= 0\, ,
\label{chap3:res2l:RQED4,3:CT:Pi2a}\\
	&\delta Z_{2b\,A}(\al_r) = \mathcal{K} \mathcal{R}' \bigg[\Pi_{2b}(q^2;\al_r) \bigg] 
	\nonum \\
	&= \mathcal{K} \bigg[ \Pi_{2b}(q^2;\al_r) \bigg] - 2\,\mathcal{K} \bigg[ \mathcal{K} \big[ \Lambda^{\mu}_{1}(\al_r)/\gamma^\mu \big]\,\Pi_{1}(q^2;\al_r) \bigg] 
	\nonum \\
	&= 0\, ,
\label{chap3:res2l:RQED4,3:CT:Pi2c}
\end{flalign}
\end{subequations}
which vanish in accordance with the fact that the singularity of each two-loop photon self-energy graph in QED$_{4,3}$ arises solely from its divergent subgraph. 
Hence, the renormalized diagrams read (for $\xi=1$):
\begin{subequations}
\label{chap3:res2l:RQED4,3:R:Pi2}
\begin{flalign}
&\Pi_{2a\,r}(q^2;\al_r) = \nonum \\ 
&=\Pi_{2a}(q^2;\al_r) - \mathcal{K} \big[ \Sigma_{1V}(\al_r) \big]\,\Pi_{1}(q^2;\al_r) - \underbrace{\delta Z_{2a\,A}(\al_r)}_{=0}
\nonum \\
&= \frac{\al^2}{\sqrt{-q^2}}\,\left( -\frac{\tilde{L}_q}{12} + \frac{1}{9} + \frac{1}{12} \right)\, ,
\label{chap3:res2l:RQED4,3:R:Pi2a}\\
&\Pi_{2b\,r}(q^2;\al_r) = 
	\nonum \\
&=\Pi_{2b}(q^2;\al_r) - 2\,\mathcal{K} \big[ \Lambda^{\mu}_{1}(\al_r)/\gamma^\mu \big]\,\Pi_{1}(q^2;\al_r) - \underbrace{\delta Z_{2b\,A}(\al_r)}_{=0}
\nonum \\
&= \frac{\al^2}{\sqrt{-q^2}}\,\left( \frac{\tilde{L}_q}{6} + \frac{\pi^2}{4} - \frac{25}{9} - \frac{1}{6}\right) \, ,
\label{chap3:res2l:RQED4,3:R:Pi2c}
\end{flalign}
\end{subequations}
where again $\tilde{L}_q = L_q + \log(4)$. Upon taking the sum of the individual contributions, Eq.~(\ref{chap3:res2l:RQED4,3:Pi2}) is straightforwardly recovered.

Finally, from the one-loop (\ref{chap3:RQED4,3:Pi1r+Sigma1r}) and two-loop (\ref{chap3:res2l:RQED4,3:Pi2}) results, the total renormalized polarization operator up to two loops can be written as:
\be
\Pi_{r}(q^2) = \Pi_{1r}(q^2)\,\left( 1 + \al_r\,\mathcal{C}^* + \Ord(\al_r^2) \right)\, , \quad \mathcal{C}^* = \frac{92-9\pi^2}{18\pi}\, ,
\label{chap3:res2l:RQED4,3:Pi-total}
\ee
where we recover the interaction correction coefficient $\mathcal{C}^*$.~\cite{Teber:2012de,Kotikov:2013kcl,Teber:2016unz}$^,$\footnote{A coefficient similar 
to $\mathcal{C}^*$ actually appeared previously in a different context: the work of Gusynin et al.\ \cite{Gusynin:2000zb}
(and, although not explicitly, in even earlier works of Gracey \cite{Gracey:1993sn} and Kotikov \cite{Kotikov:1993wr}) on $1/N$ expansion of QED$_3$. 
This is due to a mapping relating large-$N$ QED$_3$ to QED$_{4,3}$.~\cite{Kotikov:2016yrn}}
 Combining Eqs.~(\ref{chap3:res2l:RQED4,3:Pi-total}) 
and (\ref{chap2:RQED:FR:Dperp}), the transverse photon propagator up to two loops then reads:
\be
d_{r\perp}(q^2) =  \frac{\I}{2\sqrt{-q^2}}\, \frac{1}{1 + N_F\frac{\al_r \pi}{4}\,\left( 1 + \al_r \mathcal{C}^* \right)}\, ,
\label{chap3:res2l:RQED4,3:Dr:2-loop}
\ee
and essentially remains free.

\section{Two-loop fermion self-energy}
\label{Sec:Two-Loop-Sigma}

We may proceed in a similar way for the two-loop fermion self-energy:
\be
\Sigma_2(p) = \Sigma_{2a}(p) + \Sigma_{2b}(p) + \Sigma_{2c}(p) \, ,
\ee
where the diagrams are represented on Fig.~\ref{chap3:fig:two-loop:Sigma}. The latter are defined as:
\begin{subequations}
\label{chap3:def:Sigma2}
\begin{flalign}
	&-\I \Sigma_{2a}(p) = \int [\D^{d_e} k] (-\I e \gamma^\al)\,S_0(p+k)\,(-\I e \gamma^\beta)\,\times
	\nonum \\
	&\qquad \tilde{D}_{0\,\al \mu}(k)\,(\I \Pi_1^{\mu \nu}(k))\,\tilde{D}_{0\,\nu \beta}(k)\, ,
\label{chap3:def:Sigma2a} \\
	&-\I \Sigma_{2b}(p) = \int [\D^{d_e} k] (-\I e \gamma^\mu)\,S_0(p+k)\,(-\I \Sigma_1(p+k))\,\times
	\nonum \\
	&\qquad S_0(p+k)\,(-\I e \gamma^\nu)\,\tilde{D}_{0\,\mu \nu}(k)\, ,
\label{chap3:def:Sigma2b} \\
	&-\I \Sigma_{2c}(p) = \int [\D^{d_e} k] (-\I e \gamma^\mu)\,S_0(k)\,(-\I e \Lambda_1^\mu(k,p))\,\times
	\nonum \\
	&\qquad \tilde{D}_{0\,\beta \mu}(k+p)\, ,
\label{chap3:def:Sigma2c}
\end{flalign}
\end{subequations}
with the one-loop polarization operator, fermion self-energy and fermion-photon 
vertex defined in Eqs.~(\ref{chap3:def:Pi1+Sigma1+Lambda1}).

All calculations done, the general expression for the 2-loop photon self-energy diagrams of QED$_{d_\gamma, d_e}$ read:

\begin{widetext}

\begin{figure}
    \includegraphics{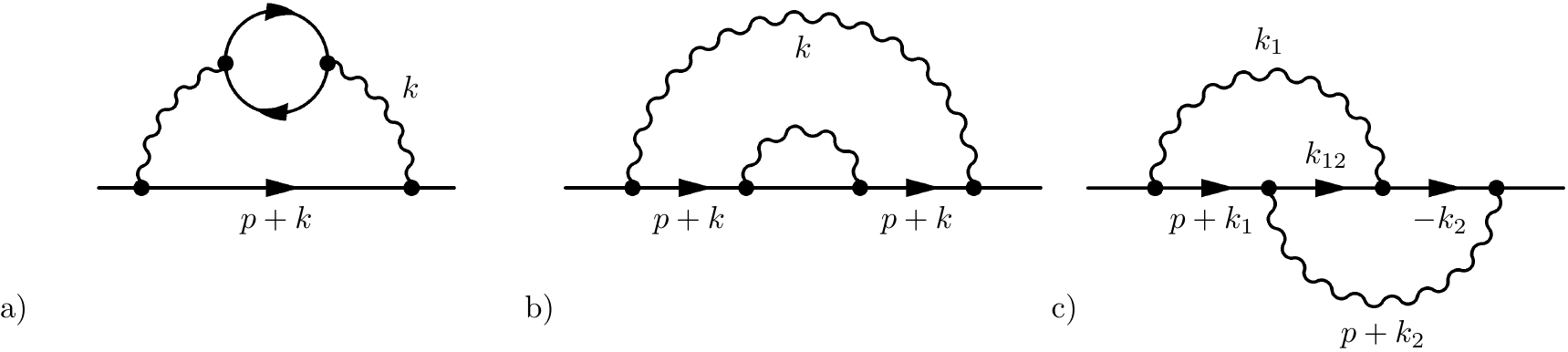}
  \caption{\label{chap3:fig:two-loop:Sigma}
  Two-loop fermion self-energy diagrams ($k_{12} = k_1-k_2$).}
\end{figure}

\begin{subequations}
\label{chap3:res:Sigma2}
\begin{flalign}
        &\Sigma_{V2a}(p^2) = 4N_F\,\bar{\al}^2 \, \left( \frac{\overline{\mu}^{\,2}}{-p^2} \right)^{2\veps_\gamma}\,\Gamma^2(1-\veps_e)\,
	\frac{(d_e-2)^2}{2(2d_\gamma-d_e-6)}\,e^{2\gamma_E \veps_\gamma}\,G(d_e,1,1)G(d_e,1,\varepsilon_\gamma-\varepsilon_e)\, ,
	\label{chap3:res:Sigma2a}\\
	&\Sigma_{V2b}(p^2) = \bar{\al}^2 \, \left( \frac{\overline{\mu}^{\,2}}{-p^2} \right)^{2\veps_\gamma}\,\Gamma^2(1-\veps_e)\,\frac{(d_e-2)(d_\gamma-3)(d_\gamma+d_e-4)}{2(d_\gamma-4)}\,\left( \xi - \frac{d_\gamma - d_e}{d_\gamma+d_e-4} \right)^2\,\times
	\nonum \\
	&\times e^{2\gamma_E \veps_\gamma}\,G(d_e,1,1-\varepsilon_e)G(d_e,1-\varepsilon_e,\varepsilon_\gamma)\, ,
	\label{chap3:res:Sigma2b}\\
	&\Sigma_{V2c}(p^2) = - \bar{\al}^2 \, \left( \frac{\overline{\mu}^{\,2}}{-p^2} \right)^{2\veps_\gamma}\,\Gamma^2(1-\veps_e)\,\frac{d_e-2}{2}\,e^{2\gamma_E \veps_\gamma}\,
	\nonum \\
	&\times \Bigg \{ \left[ d_e-4 + \frac{(d_e-2)(d_\gamma-3d_e+4)}{2(d_\gamma+d_e-4)}  -\frac{(d_\gamma+d_e-6)( d_\gamma(d_e-4)+8)}{(2d_\gamma+d_e-10)(2d_\gamma+d_e-8)}
\right . \Bigg .
\nonum \\
&\qquad -\frac{4(d_\gamma-d_e)}{d_\gamma+d_e-4} -  \frac{d_\gamma-d_e}{2d_\gamma+d_e-8}\, \left( d_e-8 - 4 \, \frac{d_\gamma+d_e-6}{d_\gamma+d_e-4} \right)
\nonum \\
&\qquad \left . -\xi\,\frac{(d_e-2)(d_\gamma-d_e)}{d_\gamma+d_e-4} +\xi^2\,\frac{d_e-2}{2} \right]\,G^2(d_e,1,1-\varepsilon_e)
\nonum \\
&+ \left[ 2d_e - d_\gamma -1  + \frac{4(d_e-2)(d_\gamma-1)}{d_\gamma+d_e-4}+\frac{8(d_\gamma-1)}{d_\gamma-4} + \frac{2(d_e-8)(d_\gamma-d_e)}{d_\gamma+d_e-6}-
\frac{4(d_\gamma-2)(d_\gamma-d_e)}{(d_\gamma-4)(d_\gamma+d_e-4)} \right .
\nonum \\
&\qquad \left . + 2\xi\,\frac{(d_\gamma-3)(d_\gamma-d_e)}{d_\gamma+d_e-4} -\xi^2 (d_\gamma-3) \right ]\,G(d_e,1,1-\varepsilon_e)G(1-\varepsilon_e,\varepsilon_\gamma)
\nonum \\
&\Bigg . - \frac{(d_\gamma-4)(d_\gamma(d_e-4)+8)}{(2d_\gamma+d_e-8)(2d_\gamma+d_e-10)}\,G(d_e,1-\varepsilon_e,1,1-\varepsilon_e,1,1) \Bigg \}\, ,
\label{chap3:res:Sigma2c-2}
\end{flalign}
\end{subequations}
where, in the last term, the complicated (UV convergent) diagram, see App.~\ref{App:MI}, comes with a factor $d_\gamma -4$ and will therefore not contribute in the case of
QED$_{4,d_e}$.

On the basis of these results, the computation of the renormalized fermion self-energies can be conveniently carried out 
using the BPHZ prescription Eq.~(\ref{chap2:def:forest}). Graphically, the renormalization constants associated with the individual 2-loop diagrams read:
\begin{subequations}
\label{chap3:def:Zpsi-forest}
\begin{flalign}
&\delta Z_{2a\,\psi}(\bar{\al}_r) = \mathcal{K}\, \bigg[
\! \! \parbox{20mm}{
    \begin{fmfgraph*}(20,20)
      \fmfleft{i1}
      \fmfright{o1}
      \fmf{phantom}{i1,i2}
      \fmf{plain,tension=1.4}{i2,i3}
      \fmf{plain}{i3,i4}
      \fmf{plain}{i4,o4}
      \fmf{plain}{o4,o3}
      \fmf{plain,tension=1.4}{o3,o2}
      \fmf{phantom}{o2,o1}
      \fmffreeze
      \fmf{phantom,left,tension=0.1,tag=2}{i2,o2}
      \fmf{phantom,left,tension=0.1,tag=3}{i3,o3}
      \fmf{phantom,left,tension=0.1,tag=4}{i4,o4}
      \fmfdot{i3,o3}
      \fmfposition
      \fmfipath{p[]}
      \fmfiset{p2}{vpath2(__i2,__o2)}
      \fmfiset{p3}{vpath3(__i3,__o3)}
      \fmfiset{p4}{vpath4(__i4,__o4)}
      \fmfi{photon}{subpath (0,3length(p3)/8) of p3}
      \fmfi{photon}{subpath (5length(p3)/8,length(p3)) of p3}
      \fmfi{plain}{point 3length(p3)/8 of p3 .. point length(p2)/2 of p2 .. point 5length(p3)/8 of p3}
      \fmfi{plain}{point 5length(p3)/8 of p3 .. point length(p4)/2 of p4 .. point 3length(p3)/8 of p3}
      \def\vert#1{%
        \fmfiv{decor.shape=circle,decor.filled=full,decor.size=2thick}{#1}}
      \vert{point 3length(p3)/8 of p3}
      \vert{point 5length(p3)/8 of p3}
   \end{fmfgraph*}
} \! \! \bigg] -  \mathcal{K}\, \bigg[~\mathcal{K}\, \bigg[~
   \parbox{15mm}{
    \begin{fmfgraph*}(15,15)
      \fmfleft{in}
      \fmfright{out}
      \fmf{boson}{in,ve}
      \fmf{plain,right,tension=0.2}{ve,vw}
      \fmf{plain,right,tension=0.2}{vw,ve}
      \fmf{boson}{vw,out}
      \fmfdot{ve,vw}
    \end{fmfgraph*}
} ~ \bigg] ~ \star ~
\parbox{15mm}{
    \begin{fmfgraph*}(15,15)
      \fmfleft{in}
      \fmfright{out}
      \fmf{plain}{in,ve}
      \fmf{plain,right,tension=0.2}{ve,vw}
      \fmf{boson,right,tension=0.2,label=$\perp$,l.s=left}{vw,ve}
      \fmf{plain}{vw,out}
      \fmfdot{ve,vw}
    \end{fmfgraph*}
}~ \bigg]\, ,
\label{chap3:deltaZ2apsi}
\\
&\delta Z_{2b\,\psi}(\bar{\al}_r,\xi_r) = \mathcal{K}\, \bigg[~
\parbox{20mm}{
    \begin{fmfgraph*}(20,20)
      \fmfleft{i1}
      \fmfright{o1}
      \fmf{plain,tension=1.7}{i1,i2}
      \fmf{plain}{i2,i3}
      \fmf{plain}{i3,o3}
      \fmf{plain}{o3,o2}
      \fmf{plain,tension=1.7}{o2,o1}
      \fmffreeze
      \fmf{photon,left,tension=0.1}{i2,o2}
      \fmf{photon,left,tension=0.1}{i3,o3}
      \fmfdot{i2,i3,o2,o3}
   \end{fmfgraph*}
} ~\bigg] -\mathcal{K}\, \bigg[~\mathcal{K}\, \bigg[~
   \parbox{15mm}{
    \begin{fmfgraph*}(15,15)
      \fmfleft{in}
      \fmfright{out}
      \fmf{plain}{in,ve}
      \fmf{plain,right,tension=0.2}{ve,vw}
      \fmf{boson,right,tension=0.2}{vw,ve}
      \fmf{plain}{vw,out}
      \fmfdot{ve,vw}
    \end{fmfgraph*}
}~ \bigg] ~ \star ~
   \parbox{15mm}{
    \begin{fmfgraph*}(15,15)
      \fmfleft{in}
      \fmfright{out}
      \fmf{plain}{in,ve}
      \fmf{plain,right,tension=0.2}{ve,vw}
      \fmf{boson,right,tension=0.2}{vw,ve}
      \fmf{plain}{vw,out}
      \fmfdot{ve,vw}
    \end{fmfgraph*}
}~ \bigg]\, ,
\label{chap3:deltaZ2bpsi}
\\
&\delta Z_{2c\,\psi}(\bar{\al}_r,\xi_r) = \mathcal{K}\, \bigg[~
\parbox{20mm}{
    \begin{fmfgraph*}(20,20)
      \fmfleft{i1}
      \fmfright{o1}
      \fmf{plain,tension=1.7}{i1,i2}
      \fmf{plain}{i2,i3}
      \fmf{plain}{i3,o3}
      \fmf{plain}{o3,o2}
      \fmf{plain,tension=1.7}{o2,o1}
      \fmffreeze
      \fmf{photon,left,tension=0.1}{i2,o3}
      \fmf{photon,right,tension=0.1}{i3,o2}
      \fmfdot{i2,i3,o2,o3}
   \end{fmfgraph*}
} \bigg] -2 \, \mathcal{K}\, \bigg[~\mathcal{K}\, \bigg[~
   \parbox{15mm}{
    \begin{fmfgraph*}(15,15)
      \fmfleft{in}
      \fmfright{e1,e2}
      \fmf{boson}{in,vi}
      \fmf{plain}{e1,v1}
      \fmf{plain,tension=0.7}{v1,vi}
      \fmf{plain,tension=0.7}{vi,v2}
      \fmf{plain}{v2,e2}
      \fmffreeze
      \fmf{boson,right,tension=0.7}{v1,v2}
      \fmfdot{vi,v1,v2}
    \end{fmfgraph*}
}~ \bigg] ~ \star ~
   \parbox{15mm}{
    \begin{fmfgraph*}(15,15)
      \fmfleft{in}
      \fmfright{out}
      \fmf{plain}{in,ve}
      \fmf{plain,right,tension=0.2}{ve,vw}
      \fmf{boson,right,tension=0.2}{vw,ve}
      \fmf{plain}{vw,out}
      \fmfdot{ve,vw}
    \end{fmfgraph*}
}~ \bigg]\, .
\label{chap3:deltaZ2cpsi}
\end{flalign}
\end{subequations}
Notice that in Eq.~(\ref{chap3:deltaZ2apsi}) the contraction of the one-loop polarization operator subgraph (which is transverse due to current conservation)
resulted in the appearance of the transverse part of the one-loop fermion self-energy Eq.~(\ref{chap3:Sigma1:T}). 
This is an example well known in the literature, see, \eg, Ref.~[\onlinecite{Kissler:2016gxn}], of the sensitivity 
of the contraction procedure to the Lorentz structure of subdiagrams. Interestingly, the transverse part is non-zero only in the reduced case ($\veps_e >0$). However, in this case
the photon self-energy is finite. It therefore vanishes in QED$_{4,d_e}$ for all values of $d_e$. Taking this into account and summing all individual contributions, 
the total 2-loop fermion renormalization constant reads:
\begin{flalign}
\delta Z_{2\,\psi}(\bar{\al}_r,\xi_r) = \mathcal{K}\, \bigg[
\! \! \parbox{20mm}{
    \begin{fmfgraph*}(20,20)
      \fmfleft{i1}
      \fmfright{o1}
      \fmf{phantom}{i1,i2}
      \fmf{plain,tension=1.4}{i2,i3}
      \fmf{plain}{i3,i4}
      \fmf{plain}{i4,o4}
      \fmf{plain}{o4,o3}
      \fmf{plain,tension=1.4}{o3,o2}
      \fmf{phantom}{o2,o1}
      \fmffreeze
      \fmf{phantom,left,tension=0.1,tag=2}{i2,o2}
      \fmf{phantom,left,tension=0.1,tag=3}{i3,o3}
      \fmf{phantom,left,tension=0.1,tag=4}{i4,o4}
      \fmfdot{i3,o3}
      \fmfposition
      \fmfipath{p[]}
      \fmfiset{p2}{vpath2(__i2,__o2)}
      \fmfiset{p3}{vpath3(__i3,__o3)}
      \fmfiset{p4}{vpath4(__i4,__o4)}
      \fmfi{photon}{subpath (0,3length(p3)/8) of p3}
      \fmfi{photon}{subpath (5length(p3)/8,length(p3)) of p3}
      \fmfi{plain}{point 3length(p3)/8 of p3 .. point length(p2)/2 of p2 .. point 5length(p3)/8 of p3}
      \fmfi{plain}{point 5length(p3)/8 of p3 .. point length(p4)/2 of p4 .. point 3length(p3)/8 of p3}
      \def\vert#1{%
        \fmfiv{decor.shape=circle,decor.filled=full,decor.size=2thick}{#1}}
      \vert{point 3length(p3)/8 of p3}
      \vert{point 5length(p3)/8 of p3}
   \end{fmfgraph*}
} \! \! \bigg] + 
\mathcal{K}\, \bigg[~
\parbox{20mm}{
    \begin{fmfgraph*}(20,20)
      \fmfleft{i1}
      \fmfright{o1}
      \fmf{plain,tension=1.7}{i1,i2}
      \fmf{plain}{i2,i3}
      \fmf{plain}{i3,o3}
      \fmf{plain}{o3,o2}
      \fmf{plain,tension=1.7}{o2,o1}
      \fmffreeze
      \fmf{photon,left,tension=0.1}{i2,o2}
      \fmf{photon,left,tension=0.1}{i3,o3}
      \fmfdot{i2,i3,o2,o3}
   \end{fmfgraph*}
} ~\bigg] +
\mathcal{K}\, \bigg[~
\parbox{20mm}{
    \begin{fmfgraph*}(20,20)
      \fmfleft{i1}
      \fmfright{o1}
      \fmf{plain,tension=1.7}{i1,i2}
      \fmf{plain}{i2,i3}
      \fmf{plain}{i3,o3}
      \fmf{plain}{o3,o2}
      \fmf{plain,tension=1.7}{o2,o1}
      \fmffreeze
      \fmf{photon,left,tension=0.1}{i2,o3}
      \fmf{photon,right,tension=0.1}{i3,o2}
      \fmfdot{i2,i3,o2,o3}
   \end{fmfgraph*}
} \bigg] 
+ \mathcal{K}\, \bigg[~\mathcal{K}\, \bigg[~
   \parbox{15mm}{
    \begin{fmfgraph*}(15,15)
      \fmfleft{in}
      \fmfright{out}
      \fmf{plain}{in,ve}
      \fmf{plain,right,tension=0.2}{ve,vw}
      \fmf{boson,right,tension=0.2}{vw,ve}
      \fmf{plain}{vw,out}
      \fmfdot{ve,vw}
    \end{fmfgraph*}
}~ \bigg] ~ \star ~
   \parbox{15mm}{
    \begin{fmfgraph*}(15,15)
      \fmfleft{in}
      \fmfright{out}
      \fmf{plain}{in,ve}
      \fmf{plain,right,tension=0.2}{ve,vw}
      \fmf{boson,right,tension=0.2}{vw,ve}
      \fmf{plain}{vw,out}
      \fmfdot{ve,vw}
    \end{fmfgraph*}
}~ \bigg]\, ,
\label{chap3:deltaZ2psi-forest}
\end{flalign}
where we have used the Ward identity Eq.~(\ref{chap2:ward:graphical}). 
Contrarily to the case of the polarization operator, Eq.~(\ref{chap3:deltaZ2psi-forest}) shows that subdivergent graphs do contribute to the renormalization of the fermion self-energy (via the
last term). 

We now apply the above formulas to specific case of QED$_{4,3}$ ($\veps_e=1/2$ and $\veps_\gamma \ra 0$).
From Eqs.~(\ref{chap3:res:Sigma2a}), (\ref{chap3:res:Sigma2b}) and Eq.~(\ref{chap3:res:Sigma2c-2}), this leads to:
\begin{subequations}
\label{chap3:res:RQED4,3:Sigma2a+b+c}
\begin{flalign}
&\Sigma_{2aV}(p^2;\bar{\al},\xi) = -\frac{2\pi^2\,N_F\, \bar{\al}^2}{3}\, \left [ \frac{1}{\veps_\gamma} - 2 L_p  + \Ord(\veps_\gamma) \right]\, ,
\label{chap3:res:RQED4,3:Sigma2a} \\
&\Sigma_{2bV}(p^2;\bar{\al},\xi) = \bar{\al}^2\, \left [ \frac{(1-3\xi)^2}{18\veps_\gamma^2} + \frac{1-3\xi}{\veps_\gamma}\,\left( \frac{11}{27} - \frac{7\xi}{9} - \frac{1-3\xi}{9}\,\tilde{L}_p \right)  + 
\frac{(1-3\xi)^2}{9}\,\left( \tilde{L}^2_p - \frac{9}{2}\,\zeta_2 \right) - \right .
\nonum \\
&\left . - \frac{2\,(11+9\xi(7\xi-6))}{27}\,\tilde{L}_p  + \frac{2\,(103 + 81\xi(7\xi-6))}{81} + \Ord(\veps_\gamma) \right]\, ,
\label{chap3:res:RQED4,3:Sigma2b} \\
&\Sigma_{2cV}(p^2;\bar{\al},\xi) = \bar{\al}^2 \left [ -\frac{(1-3\xi)^2}{9\veps_\gamma^2} + \frac{1}{\veps_\gamma}\,\left( \frac{2(1-3\xi)^2}{9}\,\tilde{L}_p  + \frac{(34-39\xi)\xi}{9} - \frac{37}{27} \right)
-\frac{2(1-3\xi)^2}{9}\,\tilde{L}_p^2 + \right .
\nonum \\
&\left . + \frac{2\,(37+3\xi(39\xi-34))}{27}\,\tilde{L}_p + \frac{71 + 21\xi(3\xi -2 )}{9}\,\zeta_2 - \frac{2(695-798\xi + 891\xi^2)}{81} + \Ord(\veps_\gamma) \right]\, ,
\label{chap3:res:RQED4,3:Sigma2c}
\end{flalign}
\end{subequations}
which are valid for an arbitrary gauge fixing parameter $\xi$. Combining Eqs.~(\ref{chap3:res:RQED4,3:Sigma2a+b+c}) with Eqs.~(\ref{chap3:def:Zpsi-forest}),
the individual counter-terms read:
\begin{subequations}
\label{chap3:res:RQED4,3:Z2a+b+c-psi}
\begin{flalign}
\delta Z_{2a\,\psi}(\bar{\al}_r) &= \mathcal{K}\, \bigg[ \Sigma_{2aV}(p^2;\bar{\al}_r) \bigg] -
\mathcal{K}\, \bigg[\underbrace{\mathcal{K}\, \bigg[\Pi_{1}(q^2;\bar{\al}_r,\xi_r) \bigg]}_{=0}\,\Sigma_{1V}^{(\perp)}(p^2;\bar{\al}_r,\xi_r) \bigg]
\nonum \\
&= -\frac{2\pi^2\,N_F\,\bar{\al}_r^2}{3\veps_\gamma}\, ,
\label{chap3:res:RQED4,3:Z2apsi} \\
\delta Z_{2b\,\psi}(\bar{\al}_r,\xi_r) &= \mathcal{K}\, \bigg[ \Sigma_{2bV}(p^2;\bar{\al}_r,\xi_r) \bigg] - \mathcal{K}\, \bigg[\mathcal{K}\, \bigg(\Sigma_{1V}(p^2;\bar{\al}_r,\xi_r) \bigg)\,\Sigma_{1V}(p^2;\bar{\al}_r,\xi_r) \bigg]
\nonum \\
&= - \frac{(1-3\xi_r)^2 \bar{\al}_r^2}{18}\,\bigg( \frac{1}{\veps_\gamma^2} - \frac{2}{3\veps_\gamma} \bigg)\, ,
\label{chap3:res:RQED4,3:Z2bpsi} \\
\delta Z_{2c\,\psi}(\bar{\al}_r,\xi_r) &= \mathcal{K}\, \bigg[ \Sigma_{2cV}(p^2;\bar{\al}_r,\xi_r) \bigg] +2\, \mathcal{K}\, \bigg[\mathcal{K}\, \bigg(\Sigma_{1V}(p^2;\bar{\al}_r,\xi_r) \bigg)\,\Sigma_{1V}(p^2;\bar{\al}_r,\xi_r) \bigg]
\nonum \\
&= \bar{\al}_r^2\,\bigg( \frac{(1-3\xi_r)^2}{9\veps_\gamma^2} - \frac{17-6\xi_r + 9 \xi_r^2}{27\veps_\gamma} \bigg)\, ,
\label{chap3:res:RQED4,3:Z2cpsi}
\end{flalign}
\end{subequations}
where, in the first line, we have used the fact that $\Pi_{1}(q^2)$ is finite in QED$_{4,3}$. The sum of Eqs.~(\ref{chap3:res:RQED4,3:Z2a+b+c-psi}) yields the total counterterm at two-loop order:
\be
\delta Z_{2\,\psi} =  -\frac{4\,\bar{\al}_r^2}{\veps_\gamma}\, \left( \zeta_2 N_F + \frac{4}{27} \right ) + \frac{(1-3\xi_r)^2\, \bar{\al}_r^2}{18\veps^2} \, .
\label{chap3:res2l:RQED4,3:Z2psi}
\ee

The individual renormalized diagrams are also straightforward to compute and read:
\begin{subequations}
\label{chap3:res:RQED4,3:Sigma2ra+b+c}
\begin{flalign}
&\Sigma_{2aVr}(p^2;\bar{\al}_r) = \Sigma_{2aV}(p^2;\bar{\al}_r) - \underbrace{\mathcal{K}\, \bigg[\Pi_{1}(q^2;\bar{\al}_r,\xi_r) \bigg]}_{=0}\,\Sigma_{1V}^{(\perp)}(p^2;\bar{\al}_r,\xi_r) 
- \delta Z_{2a\,\psi}(\bar{\al}_r)
\nonum \\
&= 8 N_F \zeta_2\,\bar{\al}_r^2\, L_p\, ,
\label{chap3:res:RQED4,3:Sigma2ar} \\
&\Sigma_{2bVr}(p^2;\bar{\al}_r,\xi_r) = \Sigma_{2bV}(p^2;\bar{\al}_r,\xi_r) - \mathcal{K} \bigg[ \Sigma_{1V}(p^2;\bar{\al}_r,\xi_r) \bigg]~\Sigma_{1V}(p^2;\bar{\al}_r,\xi_r) - \delta Z_{2b\,\psi}(\bar{\al}_r,\xi_r)
\nonum \\
&= \bar{\al}^2\, \left [ \frac{(1-3\xi_r)^2}{18}\,\left( \tilde{L}^2_p - 2\,\zeta_2\right) - \frac{4\,(1+\xi_r(6\xi_r-5))}{9}\,\tilde{L}_p  + \frac{2\,(47 -210\xi_r +243\xi_r^2)}{81} \right]\, ,
\label{chap3:res:RQED4,3:Sigma2br} \\
&\Sigma_{2cVr}(p^2;\bar{\al}_r,\xi_r) = \Sigma_{2cVr}(p^2;\bar{\al}_r,\xi_r) + 2\mathcal{K} \bigg[ \Sigma_{1V}(p^2;\bar{\al}_r,\xi_r) \bigg]~\Sigma_{1V}(p^2;\bar{\al}_r,\xi_r) - \delta Z_{2c\,\psi}(\bar{\al}_r,\xi_r)
\nonum \\
&=\bar{\al}^2\, \left [-\frac{(1-3\xi_r)^2}{9}\,\tilde{L}_p^2 + \frac{64}{9}\,\zeta_2 + \frac{2\,(3+\xi_r(7\xi_r-6))}{3}\,\tilde{L}_p + \frac{2\xi_r(82-81\xi_r)}{27} - \frac{1166}{81}  \right]\, ,
\label{chap3:res:RQED4,3:Sigma2cr}
\end{flalign}
\end{subequations}
where $\tilde{L}_p=L_p+\log 4$. The sum of Eqs.~(\ref{chap3:res:RQED4,3:Sigma2ra+b+c}) yields the total two-loop renormalized fermion self-energy:
\bea
\Sigma_{2Vr}(p^2;\bar{\al}_r,\xi_r) =&& \bar{\al}_r^2 \bigg[ 8 N_F \zeta_2\,L_p - \frac{(1-3\xi_r)^2}{18}\,\tilde{L}_p^2 + \frac{64-(1-3\xi_r)^2}{9}\,\zeta_2 +
\nonum \\
&&\qquad + \frac{2\,(7+\xi_r(9\xi_r-8))}{9}\,\tilde{L}_p - \frac{8(134 - 9\xi_r)}{81} \bigg]\, .
\label{chap3:res2l:RQED4,3:Sigma2r}
\eea

Combining the above two-loop results with the one-loop ones derived in the previous paragraphs, we recover the expression of the anomalous dimension of the fermion field 
up to two loops:~\cite{Kotikov:2013eha}$^,$\footnote{We note that using the mapping found in Ref.~[\onlinecite{Kotikov:2016yrn}] between large-$N$ QED$_3$ and QED$_{4,3}$, it is straightforward
to show that Eq.~(\ref{chap3:res2l:RQED4,3:gammapsi}) is in perfect agreement with the NLO result of Gracey for QED$_3$, Ref.~[\onlinecite{Gracey:1993sn}].}
\be
\gamma_{\psi}(\bar{\al}_r,\xi_r) = 2\bar{\al}_r\,\frac{1-3\xi_r}{3} -16\,\left( \zeta_2 N_F + \frac{4}{27} \right)\, \bar{\al}_r^2 + \Ord(\bar{\al}_r^3)\, ,
\label{chap3:res2l:RQED4,3:gammapsi}
\ee
where the two-loop contribution is gauge-invariant. Combining Eqs.~(\ref{chap3:res2l:RQED4,3:gammapsi}) and (\ref{chap3:RQED:Dyson:S}), 
the expansion of the renormalized fermion propagator up to two loops reads:
\begin{flalign}
&-\I \Sp S_r(p) = 1 +\bar{\al}_r \left( \frac{10}{9} - 2\xi_r - \frac{1-3\xi_r}{3}\,\tilde{L}_p \right) + 
\bar{\al}_r^2\,\bigg( 8 N_F \zeta_2\,L_p + \frac{(1-3\xi_r)^2}{18}\,\tilde{L}_p^2 + 
\nonum \\
&\qquad + \frac{2\,(11+3\xi_r(8-9\xi_r))}{27}\,\tilde{L}_p - \frac{4(27+\xi_r(8 - 9\xi_r))}{9} + \zeta_2\,\left(7 + \frac{2\xi_r}{3} - \xi_r^2 \right)  \bigg)
+\Ord(\bar{\al}_r^3)\, .
\label{chap3:res2l:RQED4,3:Sr}
\end{flalign}

\end{widetext}

\section{Conclusion}
\label{Sec:Conclusion}

To conclude this paper, we have provided a detailed field theoretic renormalization analysis of reduced QED up to two loops. 
The main focus was on reduced QED$_{4,3}$ (graphene at the IR Lorentz invariant fixed point) which is somehow intermediate between QED$_4$ and 
QED$_3$: it is renormalizable similarly to QED$_4$ with UV divergent fermion self-energy but has a finite photon self-energy similarly to QED$_3$.
Using the BPHZ prescription, we have provided a simple and clear renormalization of the photon and fermion self-energies. We have
straightforwardly recovered the results, previously derived via conventional renormalization, for both the interaction correction to the optical 
conductivity, Eq.~(\ref{chap3:res2l:RQED4,3:Pi-total}), and the anomalous dimension of the fermion field, Eq.~(\ref{chap3:res2l:RQED4,3:gammapsi}),
thereby lifting any possible ambiguity as to their value at the IR fixed point. This constitutes a necessary prerequisite in order to extend
our formalism to higher orders and/or to other models such as model II, Eq.~(\ref{chap1:model-inst}), that we will consider in our next investigation.~\cite{Teber:2018qcn}

\acknowledgments

The work of A.V.K. was supported in part by the Russian Foundation for 
Basic Research (Grant No.~16-02-00790-a).

\appendix

\section{Master integrals}
\label{App:MI}

In this Appendix, we recall some of the basic master integrals appearing in the main text.

The one-loop (scalar) propagator-type massless integral is defined as:
\be
J(D,p,\al,\beta) = \int \frac{[\D^D k]}{k^{2\al} (p-k)^{2\beta}}\, ,
\label{chap2:def:one-loop-p-int}
\ee
where $[\D^D k] = \D^{D} k / (2\pi)^{D}$, $p$ is the external momentum and $\al$ and $\beta$ are arbitrary indices.
In Eq.~(\ref{chap2:def:one-loop-p-int}), the momentum dependence is easily extracted from dimensional analysis which allows to 
write it the following form:
\be
J(D,p,\al,\beta) =  \frac{(p^2)^{D/2 - \al - \beta}}{(4\pi)^{D/2}}\,G(D,\al,\beta)\, ,
\label{chap2:def:one-loop-G-func}
\ee
where $G(D,\al,\beta)$ is the (dimensionless) coefficient function of the diagram: 
\be
G(D,\al,\beta) = \frac{a(\al) a(\beta)}{a(\al + \beta -D/2)}, \qquad a(\al) = \frac{\Gamma(D/2 - \al)}{\Gamma(\al)}\, .
\label{chap2:one-loop-G}
\ee

The massless two-loop propagator-type diagram is defined as:
\begin{flalign}
&J(D,p,\al_1,\al_2,\al_3,\al_4,\al_5) = 
\nonum \\
&\int \frac{[\D^Dk][\D^Dq]}{(k-p)^{2\al_1}\,(q-p)^{2\al_2}\,q^{2\al_3}\,k^{2\al_4}\,(k-q)^{2\al_5}}\, ,
\label{chap2:def:two-loop-p-int}
\end{flalign}
where the $\al_i$ ($i=1$-$5$) are five arbitrary indices. Similarly to the one-loop case, 
the momentum dependence of Eq.~(\ref{chap2:def:two-loop-p-int}) follows from dimensional analysis which allows to write this diagram in the form:
\be
J(D,p,\{\al_i\}) = \frac{(p^2)^{D-\sum_{i=1}^5 \al_i}}{(4\pi)^D}\, G(D,\{\al_i\})\, ,
\label{chap2:def:two-loop-G-func}
\ee
where $G(D,\{\al_i\})$ is the (dimensionless) coefficient function of the diagram. The expression of $G(D,1,1,1,1,\al)$ for
arbitrary $\al$ can be found in Ref.~[\onlinecite{Kotikov:1995cw}] and that of $G(D,\al,1,\beta,1,1)$ for
arbitrary $\al$ and $\beta$ can be found in Ref.~[\onlinecite{Kotikov:2013eha}].


\end{fmffile}
\end{document}